\documentclass[aps,pra,twocolumn,groupedaddress,floatfix]{revtex4-1}
\pdfoutput=1
\usepackage{graphicx}
\usepackage[latin1]{inputenc}
\usepackage{amsmath}
\usepackage{amssymb}
\usepackage{color}

\usepackage{mathtools}
\usepackage{float}
\usepackage[caption = false]{subfig}
\usepackage{hyperref}
\usepackage[T1]{fontenc}

\def\e#1{ e^{#1}}

\def\av#1{\langle#1\rangle}

\def\tder#1{\frac{d#1}{dt}}
\def\eps{\varepsilon}
\def\nup{\nu^\prime}

\def\aop{\hat{a}}
\def\adop{\hat{a}^\dagger}
\def\awop{\tilde{a}}
\def\awdop{\tilde{a}^\dagger}
\def\bop{\hat{b}}
\def\bwop{\tilde{b}}
\def\bwdop{\tilde{b}^\dagger}
\def\cop{\hat{c}}
\def\cwop{\tilde{c}}

\def\Xwop{\tilde{X}}
\def\Hop{\hat{H}}

\def\xiop{\hat{\xi}}

\def\xiwop{\tilde{\xi}}

\def\thetap{{\theta^\prime}}
\def\mnu{{-\nu}}
\def\sm{{-}}
\def\refeq#1{{\hyperref[#1]{(\ref*{#1})}}}
\def\reffig#1{{\hyperref[#1]{FIG. \ref*{#1}}}}
\def\refsec#1{{\hyperref[#1]{SEC. \ref*{#1}}}}
\def\refno#1{{\hyperref[#1]{\ref*{#1}}}}

\def\nn{\nonumber}

\newcommand{\lsz}{\left[}
\newcommand{\rsz}{\right]}
\newcommand{\lk}{\left(}
\newcommand{\rk}{\right)}
\newcommand{\lka}{\left\{}
\newcommand{\rka}{\right\}}

\begin{document}
\title{Enhanced optical squeezing from a degenerate parametric amplifier via time-delayed coherent feedback}

\author{Nikolett N\'emet}\email{nnem614@aucklanduni.ac.nz}
\author{Scott Parkins}\email{s.parkins@auckland.ac.nz}
\affiliation{Dodd-Walls Centre for Photonic and Quantum Technologies, Department of Physics, 
             University of Auckland, Private Bag 92019, Auckland, New Zealand}
\date{\today}

\begin{abstract}
A particularly simple setup is introduced to study the influence of time-delayed coherent feedback on the optical squeezing properties of the degenerate parametric amplifier (DPA). The possibility for significantly enhanced squeezing is demonstrated both on resonance and in sidebands, at a reduced pump power compared to the case without feedback. We study a broad range of operating parameters and their influence on the characteristic squeezing of the system. A classical analysis of the system dynamics reveals the connection between the feedback-modified landscape of stability and enhanced squeezing.
\end{abstract}


\maketitle

\section{Introduction}

The design and control of quantum states is central to achieving aims of modern quantum science such as high precision measurements that surpass so-called ``standard quantum limits'', and quantum computation and communication. 
A promising avenue of investigation in this respect is the emerging technique of {\it coherent quantum feedback control} (see, for example, \cite{Wisemann1994,Lloyd2000,Nelson2000,Mabuchi2008a,Gough2009,Yan2011,Iida2012,Crisafulli2013,Jacobs2014,Zhou2015}), in which a portion of the output of a system is returned to the input in such a way as to stabilize the behaviour of the system, or to steer it towards a target state. Quantum optical setups employing optical cavities provide a natural setting for such feedback control, with well-defined cavity input and output channels (through the mirrors) enabling efficient coupling of light to and from the feedback loop(s) between these channels. 

\begin{figure}[!htbp]
\centering
\includegraphics[scale=0.4]{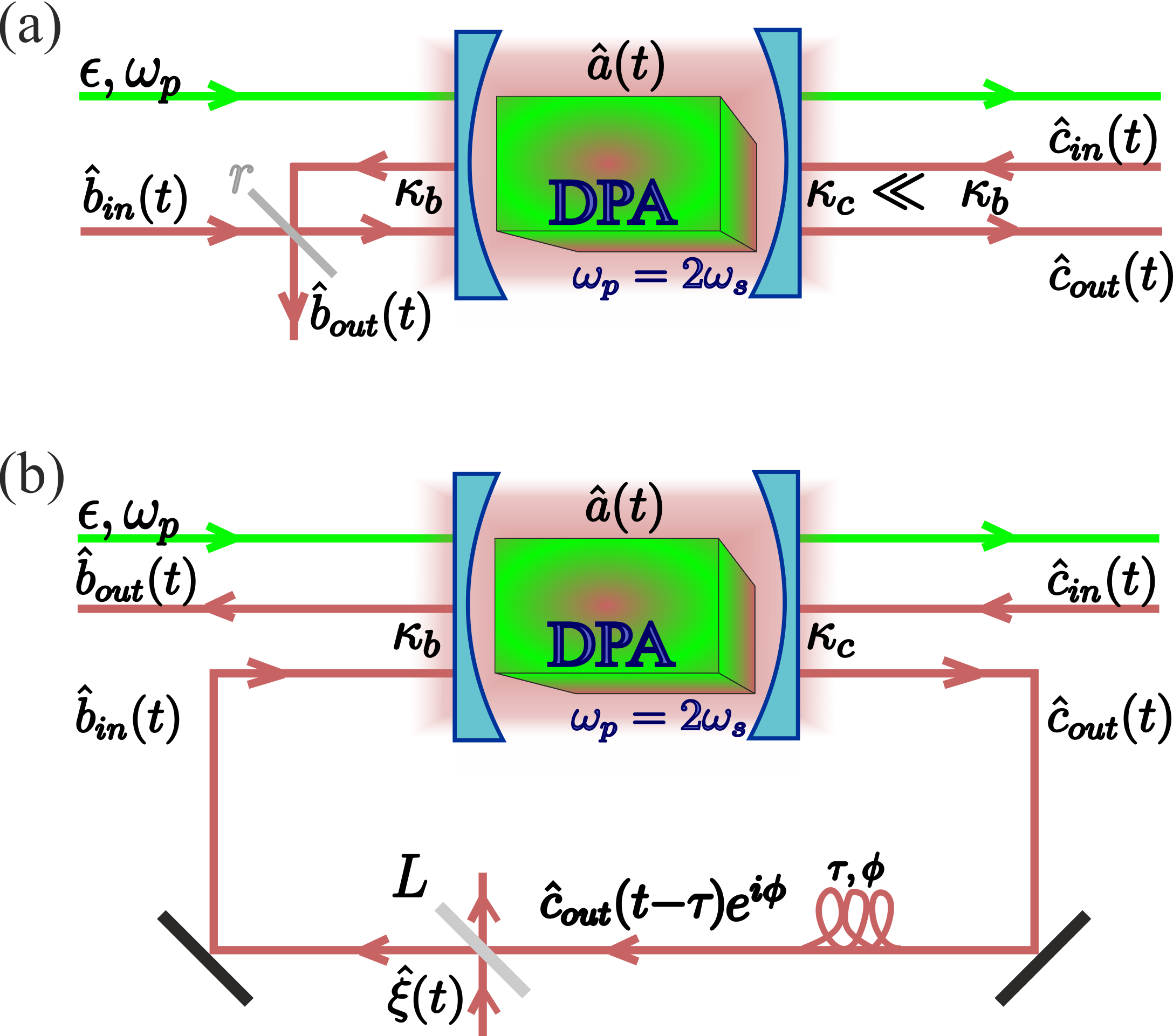}
\vspace{-.3cm}\caption{(a) Schematic of the setup proposed in \cite{Gough2009} and implemented in \cite{Iida2012} -- a DPA in a nearly one-sided cavity with feedback via a beamsplitter with reflectivity $r$. The feedback loop might include a small time delay $\tau$. (b) Schematic of our proposed setup -- a DPA in a two-sided cavity with a feedback loop that feeds the output from one mirror to the input of the other mirror. The feedback loop includes a time delay $\tau$ and an overall phase shift $\phi$, as well as loss $L$.}\label{fig:setup}
\end{figure}

Quantum squeezing of light is an important and topical example to investigate in this context. An example that has in fact been examined previously, both theoretically and experimentally, is a degenerate parametric amplifier (DPA) subject to coherent feedback via a loop containing a tunable beam splitter \cite{Gough2009,Iida2012}, as shown in Fig.~\ref{fig:setup}(a). Squeezing properties in one output port of the beamsplitter were modified by tuning the reflectivity of the beam splitter; this had the effect of varying the effective damping rate of the DPA system, leading to a modified parametric oscillation threshold and enhanced squeezing near resonance. A related scheme has been applied to a non-degenerate parametric amplifier to produce enhanced levels of continuous variable (i.e., EPR-type) entanglement between signal and idler output fields \cite{Yan2011,Zhou2015}. Meanwhile, a proposal for improving the performance of the scheme of \cite{Gough2009,Iida2012}, designed with the help of constrained nonlinear programming, has been put forward \cite{Bian2012}, which includes an additional quantum system, in particular a two-mode squeezer, as a ``controller.'' In a similar vein, a scheme that also incorporates a second optical parametric amplifier in a ``plant'' and ``controller'' coherent optical feedback configuration has been implemented and yielded enhanced squeezing at frequencies shifted away from resonance and over a broader bandwidth compared with the no-feedback system \cite{Crisafulli2013}.

None of these investigations have, however, examined the effect on the coherent feedback of a significant time-delay in the feedback loop, i.e., a time delay that is comparable to the inverse cavity field decay rate. While in some instances such a time delay might be regarded mainly as an unavoidable and possibly undesirable experimental condition, studies in the control of classical systems have highlighted unique and important applications of time-delayed feedback, such as the stabilization of unstable periodic orbits and unstable fixed points (see, for example, \cite{Pyragas1992,Hovel2005}), which has stimulated growing interest in the application of such feedback to open quantum optical systems \cite{Carmele2013,Zhang2013,Grimsmo2014,Naumann2014,Kopylov2015a,Kopylov2015b,Hein2015,Kabuss2015}. Moreover, new theoretical formalisms have also been introduced in order to deal with the difficulties of analyzing a system with time-delayed coherent feedback, which is an inherently non-Markovian problem \cite{Pichler2016,Grimsmo2015,Whalen2016,Kabuss2016}.

In this paper we address theoretically the intriguing influence of time-delayed, coherent feedback on the output squeezing spectrum of a DPA in a particularly simple configuration.
In our proposed setup, depicted in Fig.~\ref{fig:setup}(b), both mirrors of the DPA cavity are partially transmitting. Note that, while we illustrate the setup with a Fabry-P\'erot cavity configuration, the scheme can be equally well achieved, e.g., in a ring cavity configuration.
The output field on one side of the cavity is directly fed back to the input channel of the other side and homodyne detection is performed on the output field of that side. We investigate a wide variety of different operating conditions, including the effects of loss and phase shift in the feedback loop, as well as detuning of the DPA pump frequency $\omega_p$ from $2\omega_a$ (cavity resonance). We focus on the case in which the system operates below the parametric oscillation threshold, such that our model is described by linear equations of motion that can be solved exactly. 

Investigation of the output squeezing spectrum for our set up reveals significantly enhanced squeezing compared to the case without feedback, for a given pump strength. Furthermore, we show that two different types of behaviour can be distinguished depending on the value of the overall feedback phase $\phi$. Pyragas-type feedback is realized for $\phi=\pi$, which reduces the effective cavity mode damping and therefore the parametric threshold pump power required to obtain a given level of squeezing on resonance. Meanwhile, for the case in which $\phi=0$, increased time delay in the feedback loop produces enhanced squeezing in narrow sidebands displaced from resonance by a shift on the order of the cavity line width, also with a lower threshold pump power. We show with a classical analysis that the former effect is a result of a shifted threshold, corresponding to a pitchfork bifurcation of steady states, whereas the latter is a consequence of a Hopf bifurcation emerging only at long enough time delays. Comparison with the previous feedback setup of \cite{Gough2009,Iida2012} points to a clear advantage of our two-sided cavity configuration over the whole range of parametric pump strength.

The paper is structured as follows. In Section II we present the quantum mechanical model to determine the output squeezing spectrum from the left-hand side of the cavity. In Section III enhanced squeezing on resonance is demonstrated with our setup without time-delay. Section IV discusses the influence of a feedback loop with long time delay on the sideband squeezing of a DPA. The modified stability landscape as a consequence of time-delayed coherent feedback is presented in Section V. In Section VI a classical nonlinear model is constructed to study the dynamics of the system in detail. The wide range of considered parameters enables tunability in the frequency of the best squeezing, which is demonstrated in Section VII. Section VIII introduces the effects of cavity detuning and deviations of the feedback-loop phase shift on the squeezing spectra. Finally, Section IX focuses on the squeezing properties of the Pyragas-type feedback scheme that is possible with this setup in the case $\phi=\pi$.

\section{Quantum mechanical model}

\subsection{The setup}

As already mentioned in the introduction, the proposed setup consists of an optical cavity and a coherent feedback loop. As depicted in Fig.~\ref{fig:setup}(b), the cavity comprises two semitransparent mirrors with transmittivities characterized by the field decay rates $\kappa_b$ and $\kappa_c$. The nonlinear crystal inside the cavity performs parametric down-conversion on incoming pump photons of frequency $\omega_p$. For a quantum mechanical description, we use the undepleted pump approximation, whereby the pump field amplitude is incorporated in the (constant) DPA pump strength amplitude $\epsilon$. The intracavity field operator is denoted by $\aop$, whereas the input and output field operators are described by $\bop_{in/out}$ on the left-hand side and $\cop_{in/out}$ on the right-hand side, respectively. The input field on the right-hand side $(\cop_{in})$ is ordinary vacuum.

The output field on the right-hand side is fed back into the cavity through the input channel on the left-hand side \cite{Grimsmo2014}. The phase shift in the feedback loop accumulated by reflection, propagation and transmission are collected in a single phase factor $\e{i\phi}$. We allow for a time delay $\tau$ in the feedback loop. In this work, typical time delays of interest are on the order of the inverse cavity field decay rate $1/\kappa$ (where $\kappa=\kappa_b+\kappa_c$), which means that, for example, with $\kappa=2\pi\cdot10\text{ MHz}$ and $\tau=1/\kappa$, the required optical path length (in free space) would be just under 5~m. Our model accounts for losses in the feedback loop via a beam splitter, such that a portion $\sqrt{L}$ of the feedback field is lost and replaced with independent vacuum noise $\xiop(t)$. We also allow for detuning of the carrier squeezing frequency, $\omega_s=\omega_p/2$, from the cavity resonance frequency $\omega_a$.

\subsection{Operator equations of motion}

In the frame rotating at frequency $\omega_s=\omega_p/2$, the equation of motion for the intracavity field operator is given by
\begin{align}
\label{eqmot}
\tder{\aop(t)}=\ &i\lsz\Hop,\aop(t)\rsz - \kappa\aop(t) -\sqrt{2\kappa}\aop_{in}(t) \nn\\
&-\e{i\phi} k\aop(t-\tau),
\end{align}
where the Hamiltonian in the undepleted pump approximation is 
\begin{align}
\Hop=&\ \Delta\adop\aop + \frac{1}{2}i\lsz\epsilon(\adop)^2-\epsilon^*(\aop)^2\rsz ,
\end{align}
with $\Delta=\omega_a-\omega_s$, $\kappa=\kappa_b+\kappa_c$, $\epsilon=|\epsilon|\e{i\theta}$, and feedback strength
\begin{align}
\label{k}
k= 2\sqrt{\kappa_b\kappa_c(1-L)} .
\end{align}
The input field operator $\aop_{in}(t)$ takes the form 
\begin{align}
\label{ain}
\aop_{in}(t)&=\frac{1}{\sqrt{2\kappa}}\lka\sqrt{2\kappa_b}\bwop_{in}(t)+\sqrt{2\kappa_c}\cop_{in}(t)\rka ,
\end{align}
where
\begin{align}
\label{bin}
\bwop_{in}(t)&=\sqrt{1-L}\e{i\phi}\cop_{in}(t-\tau)+\sqrt{L}\xiop(t).
\end{align}

While the undepleted pump approximation significantly simplifies the problem, the two-time correlation function of input field $\aop_{in}(t)$ reflects the non-Markovian nature of the system \cite{Whalen2015}, i.e.,
\begin{align}
&\av{\aop_{in}(t)\adop_{in}(t^\prime)}=\ \delta(t-t^\prime)+ \nn\\
&~~~~~+\frac{k}{2\kappa}\lsz\delta(t-t^\prime-\tau)\e{i\phi} + \ \delta(t-t^\prime+\tau)\e{\sm i\phi}\rsz .
\end{align}

\subsection{Solution in Fourier space}

The most straightforward way of dealing with the equation of motion is to transform it into Fourier space \cite{Collett1984}. This is also a natural approach given that we wish to compute the (steady-state) spectrum of squeezing. The solution for the Fourier transform of the cavity field operator,
\begin{align}
\awop(\nu) = \frac{1}{\sqrt{2\pi}} \int_{-\infty}^\infty e^{i\nu t}\aop (t)dt ,
\end{align}
is found as
\begin{align}
\awop(\nu)=&-\frac{\sqrt{2\kappa}}{m(\nu)}\lsz d_+(\nu)\awop_{in}(\nu)+\epsilon\awdop_{in}(\mnu)\rsz,
\end{align}
where
\begin{align}
\awop_{in}(\nu)=&\frac{1}{2\sqrt{\kappa\kappa_c}}\lsz f_c(\nu)\cwop_{in}(\nu)+2\sqrt{\kappa_b\kappa_cL}\xiwop(\nu)\rsz,
\end{align}
with
\begin{align}
\label{dpm}
\quad d_\pm(\nu)&=\kappa-i(\nu\pm\Delta)+k\e{\mp i(\phi\mp\nu\tau)},\\
\label{mprime}
m(\nu)&=d_+(\nu)d_-(\nu)-|\epsilon|^2,\\
\label{fb}
f_\beta(\nu)&=2\kappa_\beta+k\e{i(\phi+\nu\tau)},\quad \beta={b,c}.
\end{align}
The output field is then obtained by applying the standard input-output formula \cite{Collett1984}:
\begin{widetext}
\begin{align}
\bwop_{out}(\nu)=\sqrt{2\kappa_b}\awop(\nu)+\bwop_{in}(\nu) 
\label{outfield}
=\ \frac{f_b(\nu)}{\sqrt{2\kappa_b}}\awop(\nu)+\sqrt{1-L}\e{i(\phi+\nu\tau)}\cwop_{in}(\nu)+\sqrt{L}\xiwop(\nu).
\end{align}

\subsection{Spectrum of squeezing}

We focus on the quantum noise properties of the output field on the left-hand side of the cavity $(\hat{b}_{out})$, as measured using optical homodyne detection. We define a generalized output quadrature operator by
\begin{align}
\Xwop_{out,\theta^\prime}(\nu) = \frac{1}{2} \left[ \e{-i\theta^\prime/2}\bwop_{out}(\nu) + \e{i\theta^\prime/2}\bwdop_{out}(-\nu) \right] ,
\end{align}
and the quadrature squeezing spectrum is then
\begin{align}
\mathcal{X}_{out,\thetap}(\nu) = \int\av{\Xwop_{out,\thetap}(\nu),\Xwop_{out,\thetap}(\nup)}d\nup .
\end{align}
From the solutions above, we obtain
\begin{align}\label{eq:chi_nu}
\mathcal{X}_{out,\thetap}(\nu) =\frac{|\epsilon|}{4\kappa_b}\frac{1}{|m(\nu)|^2} & \lka \Re{\lsz\e{i(\theta-\thetap)}\lk d_+(\nu)d_+(\mnu)+|\epsilon|^2\rk f_b(\nu)f_b(\mnu)\rsz}\right.\nn\\
&\quad\quad\quad\left.+|\epsilon|\lsz \Re{d_{\sm}(\nu)}|f_b(\mnu)|^2+\Re{d_{+}(\nu)}|f_b(\nu)|^2\rsz \vphantom{\frac{~}{~}} \right\} + \frac{1}{4} ,
\end{align}
where $\Re$ ($\Im$) denotes the real (imaginary) part of an expression.

Note that for presenting our results graphically, we use a logarithmic scale for the noise relative to the ordinary vacuum level, i.e., we plot
\begin{align}
S_{out,\thetap}(\nu)&=10\log_{10}{\lsz\frac{\mathcal{X}_{out,\thetap}(\nu)}{\mathcal{X}_{SN}(\nu)}\rsz},\nn
\end{align}
\noindent where $\mathcal{X}_{SN}(\nu)=1/4$ is the vacuum noise level.

\end{widetext}

\section{Enhancement of resonant squeezing without time delay}\label{sec:compare_theory}

In an ideal, one-sided DPA the maximum amount of squeezing, without feedback, is obtained on resonance, where the phase quadrature variance is given by \cite{Collett1984}:
\begin{align}
\label{Gard_squeez}
\mathcal{X}_{out,\theta+\pi}(0)=\frac{1}{4}\frac{(\kappa-|\epsilon|)^2}{(\kappa+|\epsilon|)^2} .
\end{align}
Here, $\theta$ is the phase of the pump amplitude $\epsilon$. Therefore, perfect squeezing is possible, in principle, when $|\epsilon|=\kappa$. In practice, technical noise and other nonlinear processes within the crystal limit the observable squeezing on resonance as the pump intensity is increased. It is in part for this reason that feedback setups are of interest.

As mentioned in the introduction, a previous proposal \cite{Gough2009,Iida2012} introduces coherent feedback to a (one-sided) DPA  as shown in Fig.~\ref{fig:setup}(a). In the ideal case ($\kappa_c=0$), the variance of the phase quadrature on resonance changes to
\begin{align}
\label{squeez_Gough}
\mathcal{X}_{out,\theta+\pi}(0)=\frac{1}{4}\left( \frac{\kappa(r)-|\epsilon|}{\kappa(r)+|\epsilon|} \right)^2 ,
\end{align}
where $r$ is the reflectivity of the beam splitter and $\kappa(r)=(1-r)\kappa/(1+r)$ is the effective decay rate of the system. Therefore, the threshold pump power can be tuned by the reflectivity of the beam splitter, which can in principle enable perfect squeezing at lower pump strengths. In practice, finite loss in the second mirror (i.e., $\kappa_c\neq 0$) limits the attainable squeezing.

In our setup, shown in Fig.~\ref{fig:setup}(b), the best squeezing without time-delay also occurs on resonance, with the case $\sin{\phi}=0$ of particular interest. When there is no loss in the feedback loop ($L=0$), we find a similar result to \refeq{squeez_Gough}, i.e.,
\begin{align}
\label{squeez_our}
\mathcal{X}_{out,\theta+\pi}(0)=\frac{1}{4}\left( \frac{\kappa+k\cos{\phi}-|\epsilon|}{\kappa+k\cos{\phi}+|\epsilon|} \right)^2 ,
\end{align}
with $\cos\phi =\pm 1$. So, when $\phi=\pi$ perfect squeezing can in principle occur at a lower pump strength determined by $\kappa-k=\kappa-2\sqrt{\kappa_b\kappa_c}$. Taken to its extreme, for a symmetric cavity ($\kappa_b=\kappa_c=\kappa/2$) this would suggest perfect squeezing even as $|\epsilon|\rightarrow 0$, but, once again, inevitable losses ($L>0$) ultimately limit the achievable squeezing. We will return to this point in more detail in Section \ref{sec:Pyragas}, when we focus on the case $\cos{\phi}=-1$ and on squeezing at resonance with Pyragas-type feedback.  

Finally, we note that, as will be shown later in Section \refno{sec:detune}, the expression above for the quadrature variance is also valid for a finite detuning $\Delta < |\epsilon|$, with the change $|\epsilon|\rightarrow |\epsilon_\Delta|=\sqrt{|\epsilon |^2-\Delta^2}$.

\begin{figure}[!htbp]
\centering
\includegraphics[width=0.8\columnwidth]{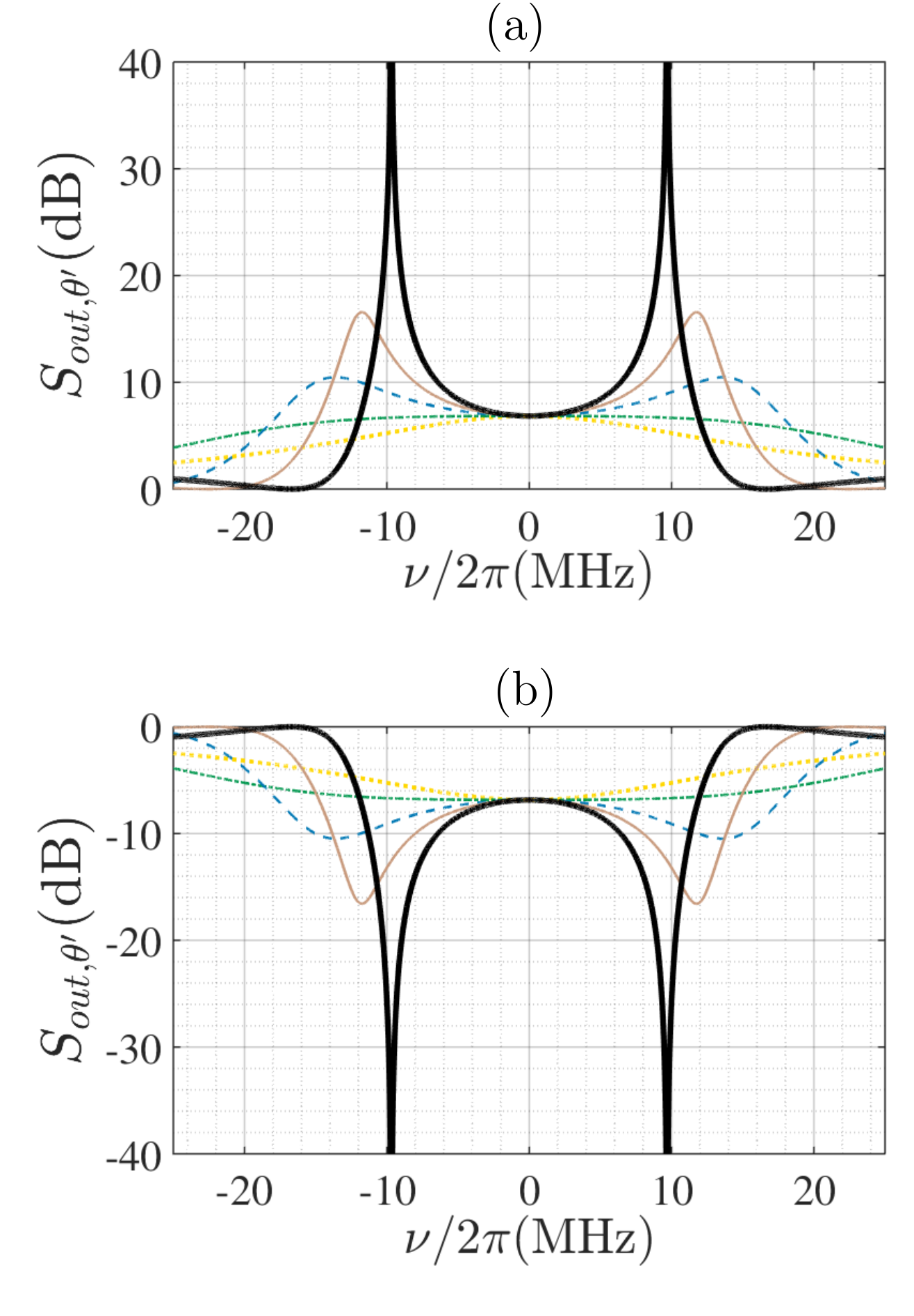}
\vspace{-.3cm}
\caption{(a) Antisqueezed ($\theta^\prime =\theta$) and (b) squeezed ($\theta^\prime =\theta +\pi$) quadrature spectra for varying time delay: $\kappa\tau=0$ (yellow dotted), $\kappa\tau=0.5$ (green dot-dashed), $\kappa\tau=1$ (blue dashed), $\kappa\tau=1.4$ (brown solid), $\kappa\tau=\kappa\tau_c=1.8833$ (thick black solid). Other parameters are $\kappa=2\pi\cdot10 \text{ MHz}$, $\kappa_b=\kappa_c=\kappa/2$, $|\epsilon|=0.75\kappa$, $\Delta=0$, $\phi=0$, and $L=0$. The characteristic frequency is $\nu_{c,|\epsilon|=0.75\kappa}=0.968\kappa$.}\label{fig:squeez_delay}
\end{figure}

\section{Enhancement of off-resonant squeezing with time delay}

With finite time delays we find that the squeezing spectra take on a more complicated structure (than for $\tau=0$), with the best squeezing moving away from resonance to sidebands that are shifted by roughly $\pm\kappa$ from resonance. Most notably, though, we find that, in the lossless case ($L=0$), perfect squeezing is in fact possible in these sidebands for a characteristic time delay, and also at a reduced pump power (cf. an equivalent one-sided DPA without feedback). A typical example of this is shown in Fig.~\ref{fig:squeez_delay}, where we consider the case of a two-sided, symmetric cavity ($\kappa_b=\kappa_c=\kappa/2$) and feedback-loop phase $\phi=0$. Note that, for a symmetric cavity such as this, the maximum squeezing possible in either output channel (i.e., through either mirror) alone, without any feedback, is 50\% (-3~dB) \cite{Collett1984}.

\subsection{Quadrature squeezing with time delay}

In the experiment performed in \cite{Iida2012}, the feedback loop gave a small time delay that was reported to have the adverse effect of narrowing the squeezing spectrum around resonance. As shown in Fig.~\ref{fig:squeez_delay}, for our feedback setup with sufficiently small time delays, the opposite is actually observed. However, with increasing time delay, a new type of behaviour arises; side-peaks appear, and, without loss in the feedback loop, we find that perfect squeezing occurs for a certain, characteristic time delay, $\tau_c$.
 
To pinpoint the characteristic time delay and frequency where perfect squeezing occurs, we note that this effect is associated with singular behaviour in the quadrature spectra (\ref{eq:chi_nu}). In particular, according to (\ref{eq:chi_nu}), divergent features are expected when $m(\nu)$ (defined in \refeq{mprime}) vanishes. Focussing in this section on the situation with $\sin\phi=0$ and $\Delta =0$, so that $d_\pm (\nu)\equiv d(\nu)=\kappa -i\nu + ke^{i\nu\tau}$, the condition $m(\nu )=0$ gives
\begin{widetext}
\begin{align}
\label{spec_tau}
\left.
\begin{array}{l} 
\Re{d(\nu)}=\kappa+k\cos{\lk\nu\tau+\phi\rk}=|\epsilon| \\ ~ \\
\Im{d(\nu)}=-\nu+k\sin{\lk\nu\tau+\phi\rk}=0
\end{array}
\right\} 
\xRightarrow{\phi=0,\pi} \left\{
\begin{array}{l}
\nu_{c}=\pm\sqrt{k^2-(\kappa-|\epsilon|)^2} \\~\\
\displaystyle
\tau_{c}=\frac{\arccos{\lk\pm\frac{|\epsilon|-\kappa}{k}\rk}}{\nu_{c}}
\end{array} \right. .
\end{align}
The two orthogonal quadrature variances, with frequency and time delay set to the characteristic values \refeq{spec_tau}, follow as
\begin{align}
\label{Sout0}
\mathcal{X}_{out,\theta}(\nu) =\ & \frac{1}{4}\frac{\lk \Re{d}(\nu)+|\epsilon|\rk^2+\Im{d}(\nu)^2-4|\epsilon|\kappa_cL}{\left|d(\nu)-|\epsilon|\right|^2} ~\xRightarrow{\nu_c,\tau_c} ~ \frac{1}{4}\frac{|\epsilon|(|\epsilon|-L\kappa_c)}{(\Re{d}(\nu_c)-|\epsilon|)^2}\rightarrow\infty ,\\
\label{Soutpi}
\mathcal{X}_{out,\theta+\pi}(\nu) =\ & \frac{1}{4}\frac{\lk \Re{d}(\nu)-|\epsilon|\rk^2+\Im{d}(\nu)^2+4|\epsilon|\kappa_cL}{\left|d(\nu)+|\epsilon|\right|^2} ~ \xRightarrow{\nu_c,\tau_c} ~ \frac{1}{4}\frac{L\kappa_c}{|\epsilon|} \xrightarrow{L\rightarrow0}\ 0.
\end{align}
\end{widetext}
This shows explicitly that without loss in the feedback loop perfect squeezing can be obtained for a given driving strength at a frequency shifted from resonance, when the time-delay is set to \refeq{spec_tau}. With finite loss the degree of squeezing is obviously reduced, but as we shall see below, significant improvements over a conventional, one-sided DPA are still possible.

Before continuing, we note that the expressions in (\ref{spec_tau}), which are specific to the cases $\phi=\{ 0,\pi\}$, are only valid in certain regimes of parameter space that differ somewhat for each case. These regimes will be elucidated in detail in the stability analysis of Section~\ref{sec:stability}.

\subsection{Comparison with no-feedback DPA \cite{Collett1984}}

\begin{figure}[!htbp]
\centering
\includegraphics[width=0.37\textwidth]{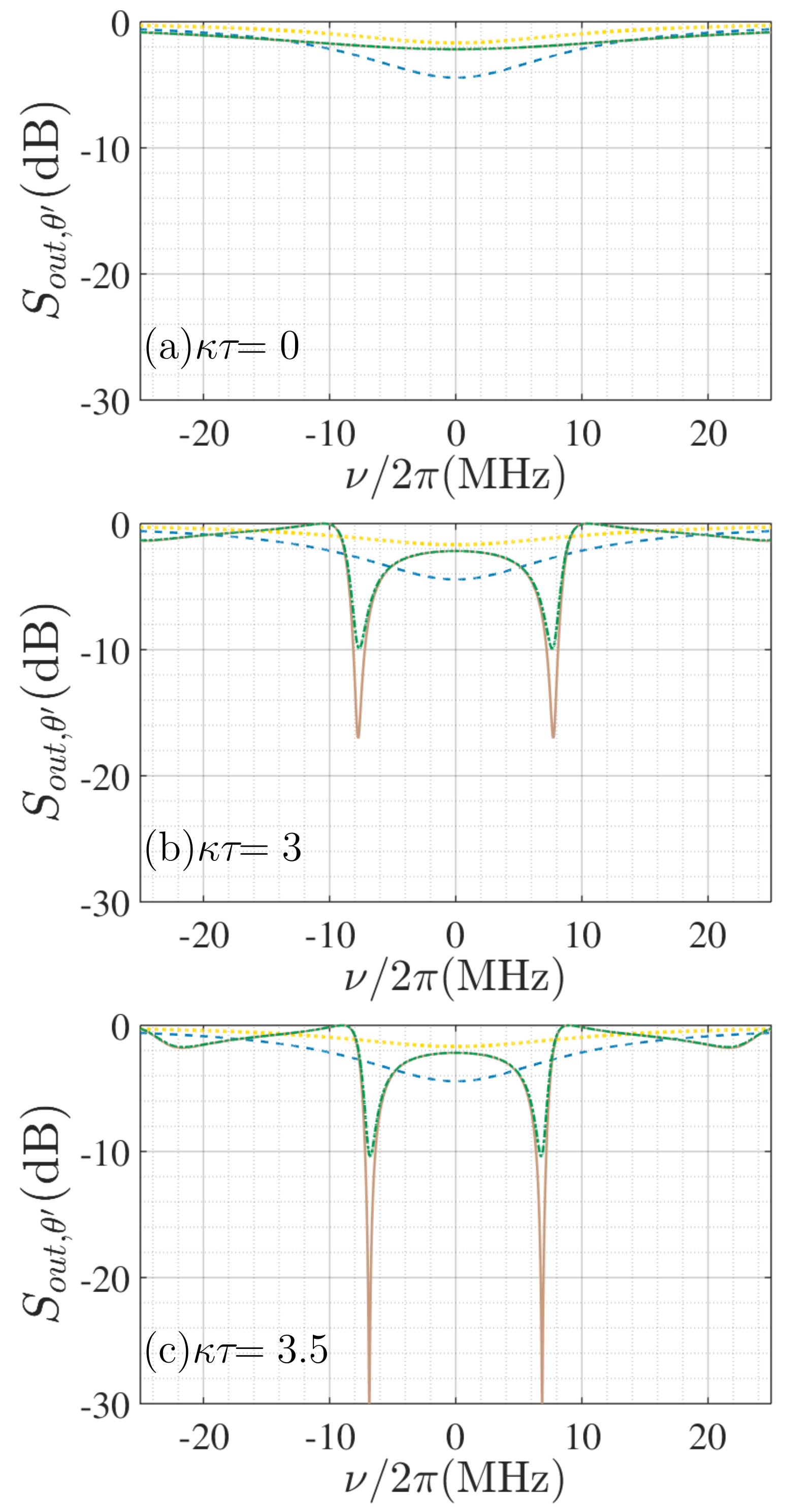}
\vspace{-.3cm}\caption{Squeezing spectra for $\thetap=\theta+\pi$ with varying time delays and loss. Our results with feedback are shown for $L=0$ (brown solid) and $L=0.05$ (green dot-dashed). Other parameters are $\kappa=2\pi\cdot10 \text{ MHz}$, $\kappa_b=\kappa_c=\kappa/2$ (symmetric cavity), $|\epsilon|=\kappa/4$, $\Delta=0$, and $\phi=0$. The characteristic time delay and frequency are $\kappa\tau_c=3.657$ and $\nu_c=0.661\kappa$. Corresponding theoretical results without feedback \cite{Collett1984} are plotted for the symmetric cavity (yellow dotted) and for a one-sided cavity ($\kappa_c=0$, $\kappa_b=\kappa$) (blue dashed). }\label{fig:squeez3}
\end{figure}

\begin{figure}[!htbp]
\centering
\includegraphics[width=0.37\textwidth]{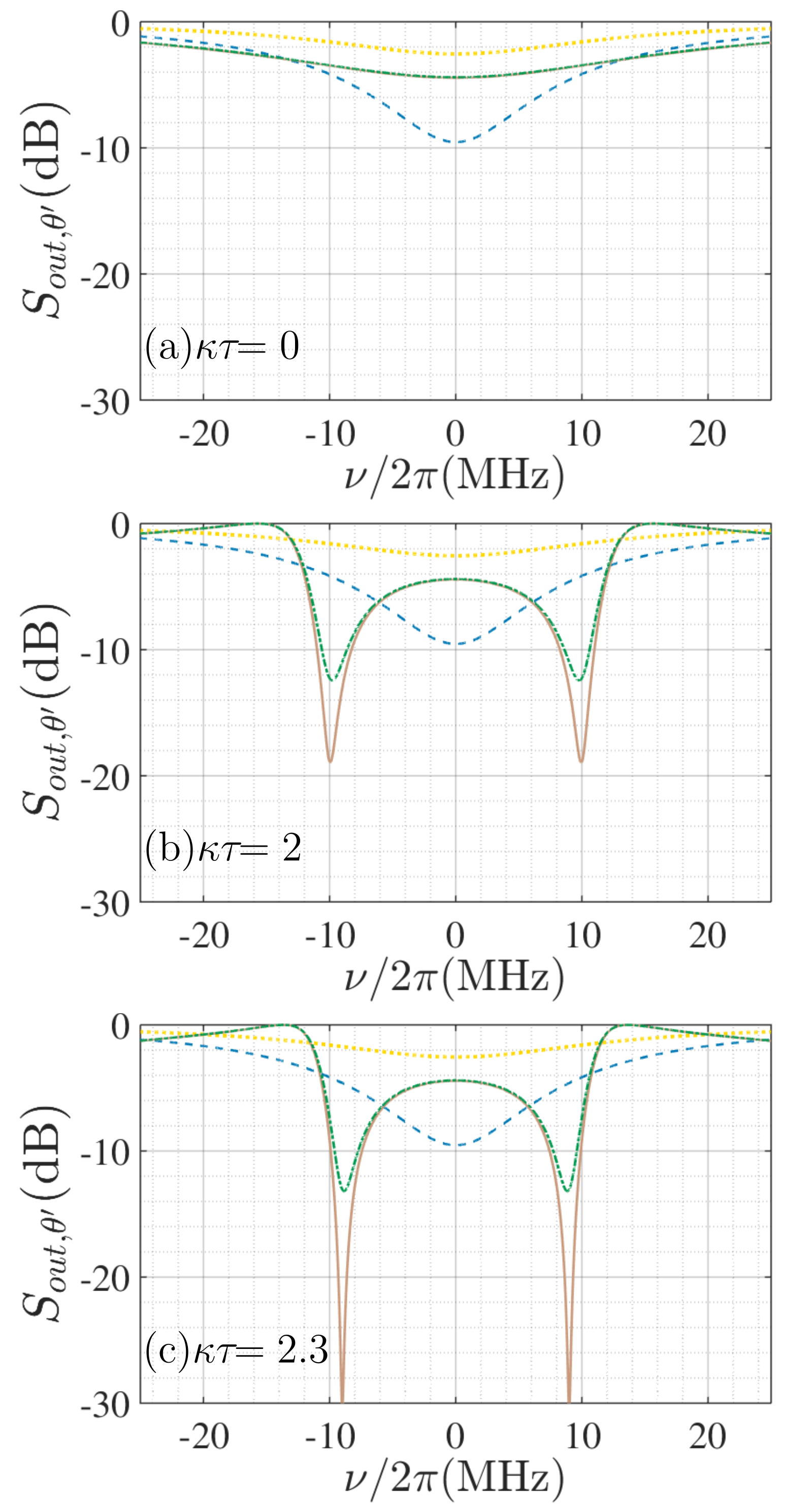}
\vspace{-.3cm}
\caption{Squeezing spectra for $\thetap=\theta+\pi$ with varying time delays and loss. Our results with feedback are shown for $L=0$ (brown solid) and $L=0.05$ (green dot-dashed). Other parameters are $\kappa=2\pi\cdot10 \text{ MHz}$, $\kappa_b=\kappa_c=\kappa/2$ (symmetric cavity), $|\epsilon|=\kappa/2$, $\Delta=0$, and $\phi=0$. The characteristic time delay and frequency are $\kappa\tau_c=2.418$ and $\nu_c=0.866\kappa$. Corresponding theoretical results without feedback \cite{Collett1984} are plotted for the symmetric cavity (yellow dotted) and for a one-sided cavity ($\kappa_c=0$, $\kappa_b=\kappa$) (blue dashed). }\label{fig:squeez4}
\end{figure}

First of all, we consider how the output squeezing from our setup performs in comparison with the case of no feedback. We consider here, in particular, the situation in which the phase shift $\phi=0$ (which, we note, corresponds to the destructive interference case of \cite{Crisafulli2013}). In Figs.~\ref{fig:squeez3} and \ref{fig:squeez4} we plot squeezing spectra for several feedback time delays in the case of a symmetric cavity ($\kappa_b=\kappa_c=\kappa/2$) and for two different values of the pump strength, $|\epsilon|=\kappa/4$ (Fig.~\ref{fig:squeez3}) and $|\epsilon|=\kappa/2$ (Fig.~\ref{fig:squeez4}). The squeezing spectra are also plotted for a DPA without feedback in both the symmetric and one-sided cavity configurations, with the same values of $\kappa$ and $|\epsilon|$.

For our feedback scheme with zero time delay (Figs.~\ref{fig:squeez3}(a) and \ref{fig:squeez4}(a)), we already find squeezing levels that are higher than the equivalent symmetric cavity case without feedback. The best squeezing does not, however, exceed that of the equivalent one-sided-cavity DPA without feedback, until the time delay is sufficiently long. When $\tau$ is large enough, squeezing around the characteristic frequency $\nu_c$ can far exceed that possible for the one-sided-cavity DPA without feedback at the equivalent pump strength, as demonstrated in Figs.~\ref{fig:squeez3}(b,c) and \ref{fig:squeez4}(b,c). The comparison with the symmetric-cavity DPA without feedback is even more dramatic, given that, as mentioned earlier, in this case the maximum possible squeezing in the output from one mirror is limited to 50\% ($-3$~dB) at resonance as $|\epsilon|\rightarrow\kappa$ \cite{Collett1984}. 

\begin{figure}[!htbp]
\centering
\includegraphics[width=0.36\textwidth]{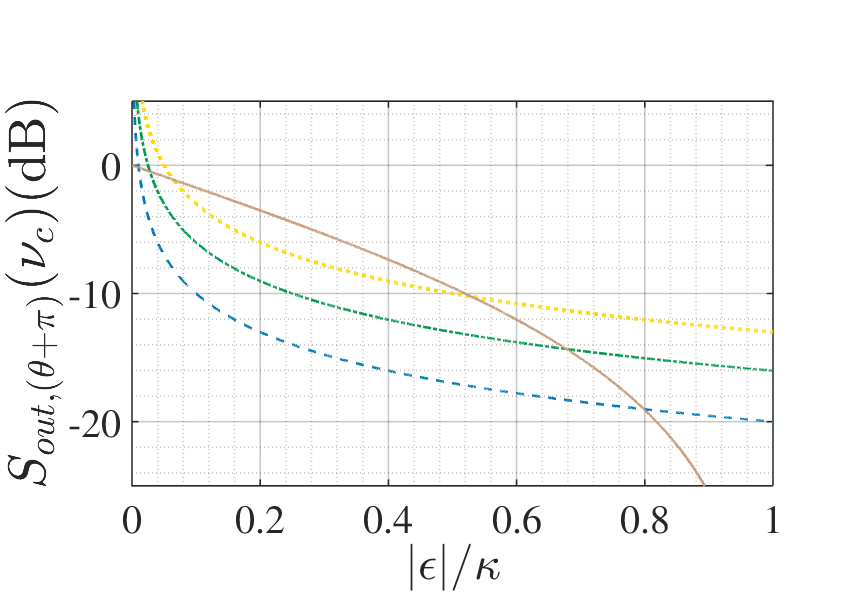}
\vspace{-.3cm}
\caption{Maximum squeezing obtainable with our feedback setup at loss levels $L=0.02$ (blue dashed), $L=0.05$ (green dot-dashed), $L=0.10$ (yellow dotted), compared with the theoretical limit of squeezing on resonance from a DPA in a one-sided cavity ($\kappa_b=\kappa$, $\kappa_c=0$) without feedback (brown-solid). Parameters for our scheme are $\kappa_b=\kappa_c=\kappa/2$, $\Delta=0$, $\phi=0$, $\nu=\nu_c$, and $\tau=\tau_c$.}\label{fig:deep_squeez}
\end{figure}

With finite loss $L>0$, the degree of squeezing is, of course, reduced. Note that loss changes the feedback strength, $k=2\sqrt{\kappa_b\kappa_c(1-L)}$, and therefore the characteristic frequency and time-delay \refeq{spec_tau}. Nevertheless, the feedback scheme can still provide significant enhancement over a broad range of parameters. 
This is highlighted in Figure \refno{fig:deep_squeez}, where the feedback limit for $\mathcal{X}_{out,\theta+\pi}(\nu_c)$ given in \refeq{Soutpi} is compared, as a function of pump strength $|\epsilon|$, with the result \refeq{Gard_squeez} for $\mathcal{X}_{out,\theta+\pi}(0)$ for a DPA in an ideal one-sided cavity. This figure also highlights that the maximum possible degree of enhancement becomes more sensitive to loss with increasing pump strength. 

\enlargethispage{1cm}

\vspace{-3cm}

\newpage
\subsection{Comparison with beamsplitter feedback \cite{Gough2009,Iida2012}}

\begin{widetext}

\begin{figure}[!htbp]
\centering
\includegraphics[width=0.9\textwidth]{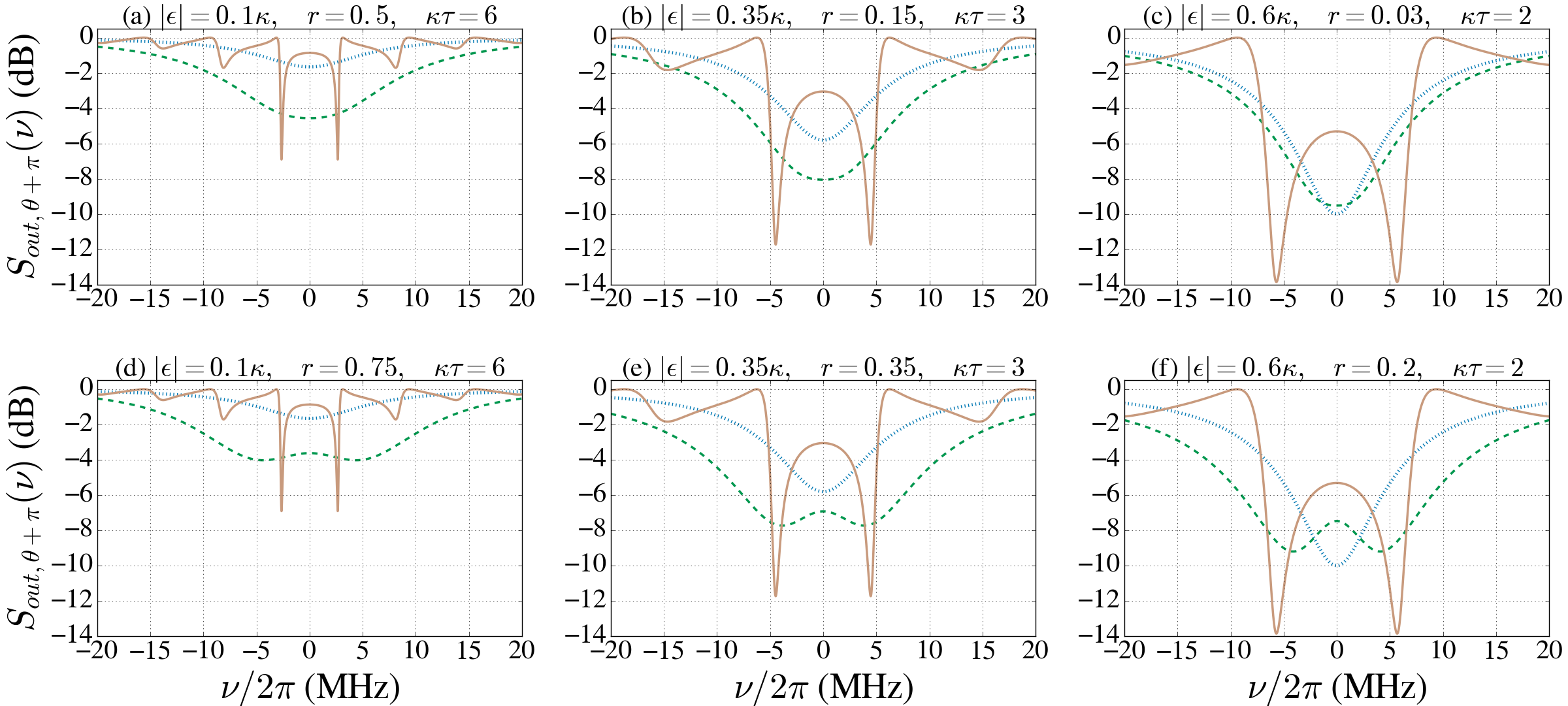}
\vspace{-.3cm}
\caption{Squeezing spectra for varying pump strengths $|\epsilon|$ as indicated. The result without feedback \cite{Collett1984} and the result with feedback of the form implemented in \cite{Iida2012} are plotted for an almost one-sided cavity ($\kappa_c=0.005\kappa$) as blue-dotted and green-dashed curves, respectively.  Our results with feedback are shown as brown solid lines. The parameter $r$ denotes the reflectivity of the beam splitter in the feedback setup of \cite{Gough2009,Iida2012}. Parameters for our scheme are $\thetap=\theta+\pi$, $\kappa=2\pi\cdot10 \text{ MHz}$, $\kappa_b=\kappa_c=\kappa/2$, $\Delta=0$, $\phi=0$, $L=0.05$.}\label{fig:furu_compare}
\end{figure}

\end{widetext}

Next, in Fig.~\refno{fig:furu_compare}, we compare our results with those for the coherent feedback setup of \cite{Gough2009,Iida2012}. This setup proves to be inefficient above $|\epsilon|=0.6\kappa$ \cite{Iida2012}, while our setup produces an enhanced level of squeezing, even in the presence of loss, at this driving strength. Note that, in the second row of Fig.~\refno{fig:furu_compare}, for higher values of reflectivity $r$ in the setup of \cite{Gough2009,Iida2012}, higher sideband squeezing is possible because of a small amount of time delay that we take into account in computing the (dashed) curves in the figure. However, the sideband squeezing does not exceed the maximum obtained on resonance for lower values of $r$.

\section{Stability analysis}\label{sec:stability}

Perfect squeezing can be associated with a change in the stability of the system, which we examine now using an eigenvalue analysis of the equations of motion for the average quadrature amplitudes, 
\begin{align}
X_1(t) &= \frac{1}{2} \lk \e{-i\theta/2}\av{\aop(t)} + \e{i\theta/2}\av{\adop(t)} \rk,
\end{align}

\begin{align}
X_2(t) &= \frac{1}{2i} \lk \e{-i\theta/2}\av{\aop(t)} - \e{i\theta/2}\av{\adop(t)} \rk ,
\end{align}
which are linear in the undepleted pump approximation. An exponential ansatz for the time evolution of $X_{1,2}(t)$ gives the following characteristic equation:
\begin{widetext}
\begin{align}
\label{stab_mat}
\lk\begin{array}{ccc} \lambda-|\epsilon|+\kappa+ k\cos{\phi}\ \e{-\lambda\tau} & -(\Delta+k\sin{\phi}\ \e{-\lambda\tau})\\\Delta+k\sin{\phi}\ \e{-\lambda\tau} & \lambda+|\epsilon|+\kappa+ k\cos{\phi}\ \e{-\lambda\tau}\end{array}\rk
\lk\begin{array}{ccc} X_1(t)\\X_2(t)\end{array}\rk = 0.
\end{align}
\end{widetext}

Let us focus on the same special case as previously, i.e., $\sin{\phi}=0$ and $\Delta=0$. With these choices, the following equations can be derived for the real and imaginary parts of the eigenvalues, $\lambda_r$ and $\lambda_s$:
\begin{align}
\label{real}
\lambda_r+\kappa-|\epsilon| &= - k\cos{(\phi)}\cos{(\lambda_i\tau)}\e{-\lambda_r\tau},\\
\label{imag}
\lambda_i &=  k\cos{(\phi)}\sin{(\lambda_i\tau)}\e{-\lambda_r\tau}.
\end{align}
We note here that when there is a finite detuning $\Delta$ between the cavity resonance and the carrier frequency, the optimal orthogonal quadratures are simply rotated (see Section~\refno{sec:detune}).

\subsection{$\phi=0$}

In zeroth order the system is stable as long as
\begin{align}
\kappa+k\cos{\lk \nu\tau\rk}> |\epsilon|.
\end{align}
Therefore, in this case and with finite time delay, off-resonant components become unstable sooner than the resonant component. Eqs.~\refeq{spec_tau} are special cases of \refeq{real} and \refeq{imag}, where the real part of the eigenvalue vanishes, i.e., they correspond to the change in stability of the system with increasing time-delay.

\subsection{$\phi=\pi$} \label{sec:stab_pyrag}

A phase shift $\phi=\pi$ results in a reduced effective decay rate, as evidenced by the stability condition
\begin{align}
\kappa-k>|\epsilon|,
\end{align}
which in turn means a reduced range of pump power (for below-threshold behaviour). It is interesting to note that in the plant-controller setup of \cite{Crisafulli2013}, reduced stability was indeed observed in the corresponding (constructive interference) case.

We note also that Eqs.~\refeq{spec_tau} as the conditions for emerging side-peaks suggest a regime that is already unstable ($\cos{(\nu_c\tau_c)}<1$ unless $\nu_c=0$ or $\tau_c=0$). Hence, for $\phi=0$ the best squeezing always occurs on resonance.

\section{Classical nonlinear model}

In the quantum model of the degenerate parametric amplifier without feedback, perfect squeezing with an ideal, one-sided cavity coincides with the change of stability at the oscillation threshold $(|\epsilon|=\kappa)$ \cite{Collett1984}.
As the complete model with feedback, including the above-threshold regime, is a nonlinear, non-Markovian problem, the quantum mechanical treatment is far from trivial. Although proposals for theoretical analyses of open quantum systems with delayed feedback have appeared recently \cite{Pichler2016,Grimsmo2015,Whalen2016,Kabuss2016}, all of them have their own computational difficulties. In our case, a classical approach, where the pump mode dynamics is included and pump depletion is allowed for, still provides a useful and relevant study of the basic dynamical properties.

By taking the expectation value of \refeq{eqmot} and using the equation of motion for the pump mode derived in \cite{Carmichael2009}, the following classical equations can be obtained (for $\Delta=0$),
\begin{align}
\label{nonlin_sub}
\dot{\bar{\eps}}(t)&=-\kappa\bar{\eps}(t)+\kappa\bar{\eps}^*(t)\bar{\eps}_p(t)
-\e{i\phi}k\bar{\eps}(t-\tau) ,\\
\label{nonlin_pump}
\dot{\bar{\eps}}_p(t)&=-\kappa_p\lk\bar{\eps}_p(t)+\bar{\eps}^2(t)-x\rk ,
\end{align}
where $\bar{\eps}$ ($\bar{\eps}_p$) is the signal (pump) amplitude, and the driving constant $x$ corresponds to the previously defined constant $|\epsilon|$ in units of $\kappa_p$. Numerical investigation of this system of equations is performed by using the Matlab package DDE-BIFTOOL and the dde23 solver module.

\begin{figure}[!htbp]
\includegraphics[width=0.4\textwidth]{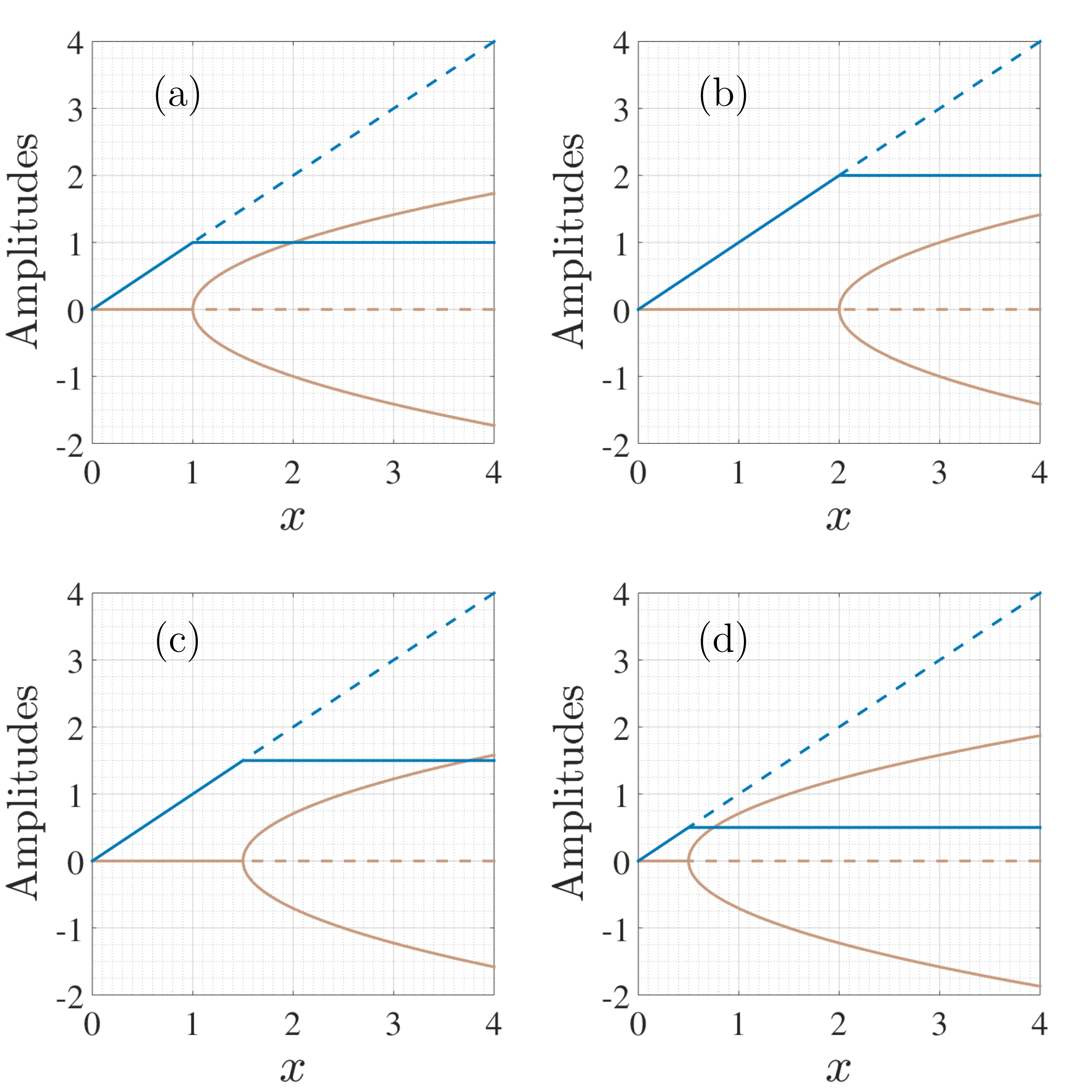} 
\vspace{-.3cm}
\caption{Steady state solutions of Eqs.~\refeq{nonlin_sub} and \refeq{nonlin_pump} for $\bar{\eps}$ (brown) and $\bar{\eps}_p$ (blue) (a) without feedback ($k=0$), and with instantaneous ($\tau=0$) feedback: (b) $k=\kappa$, $\phi=0$, (c) $k=\kappa/2$, $\phi=0$, (d) $k=\kappa/2$, $\phi=\pi$ (Pyragas-type feedback -- see Section \ref{sec:Pyragas}). Stable solutions are shown as solid lines, unstable solutions as dashed lines. We set $\kappa_p=\kappa$.}\label{fig:bifurc}
\end{figure}

The bifurcation diagram in Fig.~\ref{fig:bifurc}(a) demonstrates the threshold in a DPA without feedback ($k=0$). When the normalized pump strength $x$ reaches $1$, the previously stable solution becomes unstable and two new stable equilibrium solutions for $\bar{\eps}$ appear. Detailed discussion of this system can be found in \cite{Carmichael2009}.

\subsection{Feedback without time delay}

With instantaneous feedback, i.e., $k\neq 0$ and $\tau=0$, steady state solutions of \refeq{nonlin_sub} and \refeq{nonlin_pump} are, below threshold,
\begin{align}
\bar{\eps}_{SS}=0, \qquad \bar{\eps}_{p,SS}=x,
\end{align}
on threshold,
\begin{align}
\label{threshold}
\bar{\eps}_{SS}=0, \qquad \bar{\eps}_{p,SS}=x_{th}=\left|1+\frac{k}{\kappa}\e{i\phi}\right|,
\end{align}
and, above threshold,
\begin{align}
\bar{\eps}_{SS}&=\pm\sqrt{\frac{\zeta}{2}}\lsz\sqrt{1+\frac{\xi}{x}}-i\sqrt{1-\frac{\xi}{x}}\rsz ,
\\
\bar{\eps}_{p,SS}&=\frac{1}{x}\lsz\lk x^2-\xi\zeta\rk+i\zeta\frac{k}{\kappa}\sin{\phi}\rsz,
\end{align}
where
\begin{align}
\xi=\sqrt{x^2-\lk\frac{k}{\kappa}\sin{\phi}\rk^2}
\end{align}
and
\begin{align}
\zeta=\xi-\lk1+\frac{k}{\kappa}\cos{\phi}\rk .
\end{align}

Feeding back the output into the input channel of the system with different coupling strengths $k$ results in a shift of the oscillation threshold \refeq{threshold} (Figs.~\ref{fig:bifurc}(b) and \ref{fig:bifurc}(c)). This shift implies a different parametric pump strength where, theoretically, perfect squeezing can occur \refeq{squeez_our}. Hence, as already mentioned in Section \ref{sec:compare_theory}, without time delay, our setup is similar to the case of \cite{Gough2009,Iida2012}.

With the choice of feedback phase $\phi=0$ the threshold is shifted to higher driving strengths with increasing feedback strength, while at $\phi=\pi$ it shifts in the opposite direction (Fig.~\ref{fig:bifurc}(d)). Because of the non-trivial mixing of quadrature amplitudes \refeq{stab_mat}, in the case of a phase shift such that $\sin{\phi}\neq0$, complex values of the amplitudes emerge.

\subsection{Feedback with time delay}

As seen previously \cite{Grimsmo2014,Carmele2013,Naumann2014}, the introduction of a long enough time delay to the feedback changes the stability of the steady state solutions. 
At the point where the stability changes as a result of a finite time delay, a pair of imaginary stability eigenvalues appears. 
This process is called Hopf bifurcation and, in the following, the imaginary eigenvalues will be referred to as the Hopf frequencies $(\omega_{\rm Hopf})$. 

\begin{figure}[!htbp]
\centering
\includegraphics[width=0.4\textwidth]{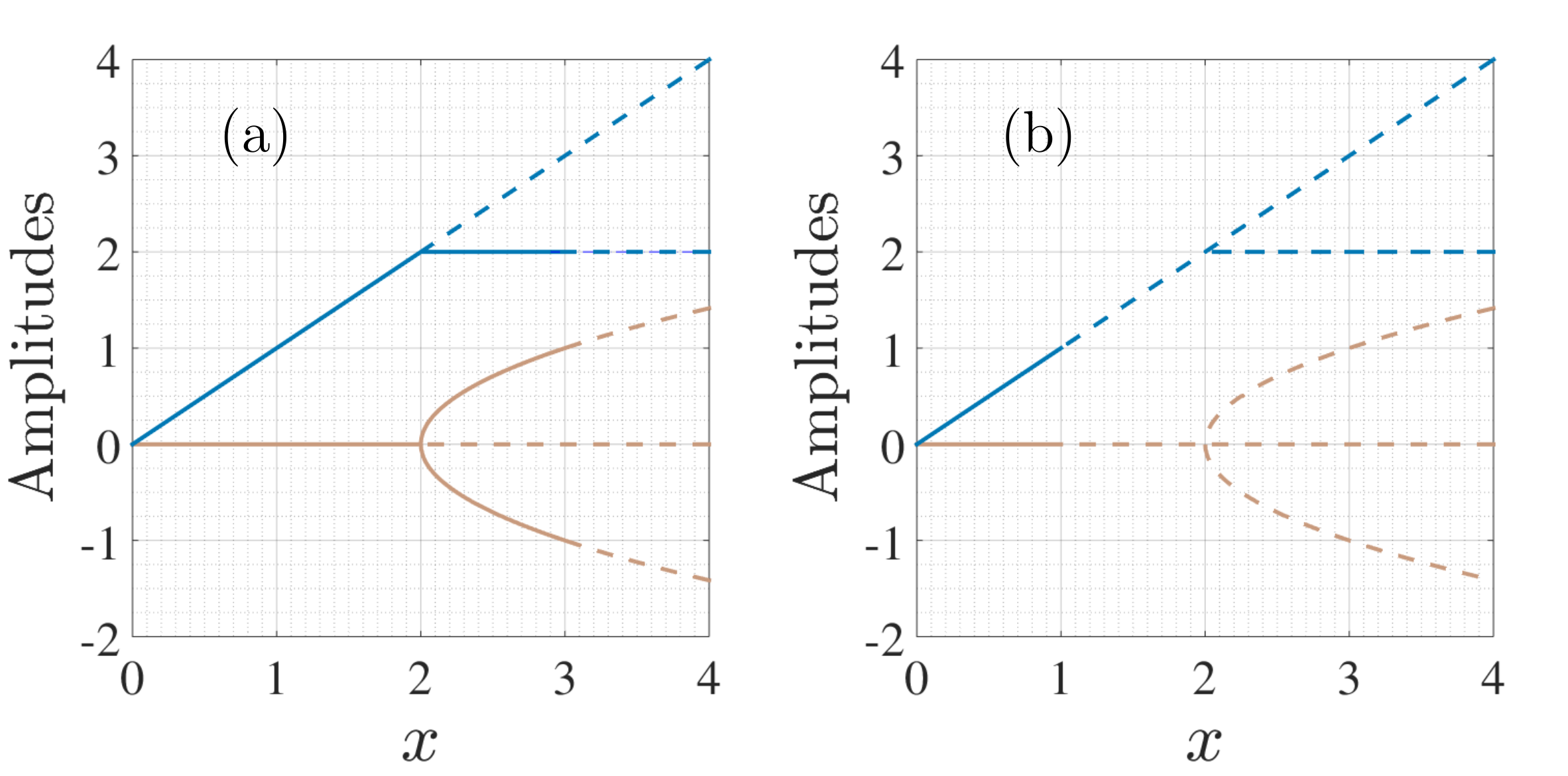}
\vspace{-.3cm}
\caption{Steady state solutions of Eqs.~\refeq{nonlin_sub} and \refeq{nonlin_pump} for $\bar{\eps}$ (brown) and $\bar{\eps}_p$ (blue) with feedback ($k=\kappa$, $\phi=0$) and time delay (a) $\kappa\tau=0.6$, (b) $\kappa\tau=1.57$. Stable solutions are shown as solid lines, unstable solutions as dashed lines. We set $\kappa_p=\kappa$.}\label{fig:Bifurc_feedback}
\end{figure}

For driving strengths less than the original threshold (e.g., Fig.~\ref{fig:Bifurc_feedback}), below the Hopf point the system converges to the steady states, whereas above it starts to oscillate with a period $2\pi/\omega_{\rm Hopf}$ (Fig.~\ref{fig:Oscillation}).
These stable oscillations were interpreted as trapped modes in \cite{Tabak2015a}. For Hopf bifurcations at larger driving strengths, above the original threshold (e.g. Fig.~\ref{fig:Bifurc_feedback}(a)) periodic solutions become unstable with increasing pump strength.  We find that, in our system, lack of stability in the branch below threshold also means de-stabilization of the steady state solutions above threshold (Fig.~\ref{fig:Bifurc_feedback}(b)).

\begin{widetext}
~
\vspace{-.8cm}
\begin{figure}[!htbp]
\includegraphics[width=0.85\textwidth]{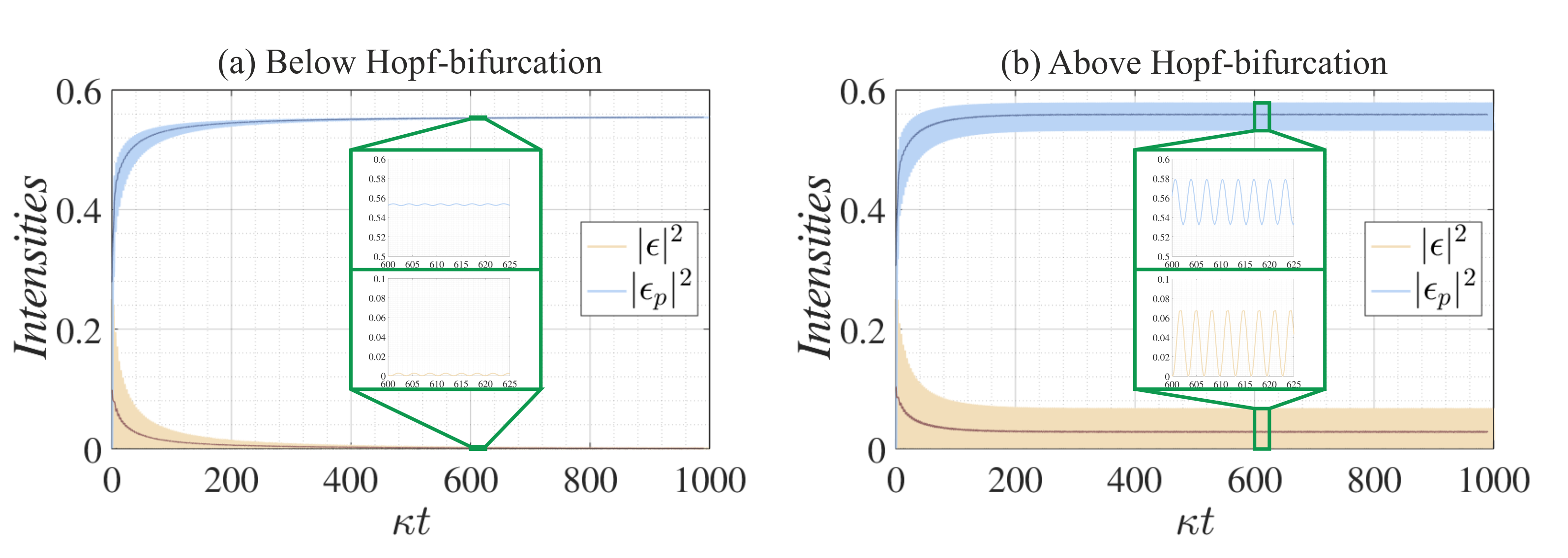}
\vspace{-.3cm}
\caption{Time evolution of the system of equations \refeq{nonlin_sub} and \refeq{nonlin_pump} with initial values $\epsilon(0)=0.5$ and $\epsilon_p(0)=0$ and (a) $x=0.745$, (b) $x=0.78$. Other parameters are $\kappa_p=\kappa$, $k=\kappa$, $\kappa\tau=1.8833$, $\phi=0$.}\label{fig:Oscillation}
\end{figure}

\end{widetext}

\subsection{Tunability of Hopf bifurcation}

An example of where the branch of Hopf bifurcations is situated in parameter space can be seen in Fig.~\ref{fig:ddetuning}(a).
Changes in the Hopf frequency are also shown in terms of the parametric pump strength and the time delay in Figs.~\ref{fig:ddetuning}(b) and \ref{fig:ddetuning}(c). The curve in Fig.~\ref{fig:ddetuning}(b) takes the following form below the original threshold,
\begin{align}
\lk\frac{\omega_{\rm Hopf}}{\kappa}\rk^2+(x-1)^2=1,
\end{align}
which for $k/\kappa=1$ and $\kappa x=|\epsilon|$ gives back the result in \refeq{spec_tau}.

Note that previously, without pump depletion, perfect squeezing occurred when the real part of the stability eigenvalue disappeared and the imaginary part corresponded to the characteristic frequency in \refeq{spec_tau}. Now, Hopf bifurcation arises in the same fashion at the same values of time delay. These characteristics imply that the sharp peaks in Figs.~\ref{fig:squeez3} and \ref{fig:squeez4} correspond to the internal dynamics of the system associated with the Hopf frequency. When pump depletion is considered, according to the classical model, much higher values of the characteristic frequency might be expected.

\begin{widetext}
~

\vspace{-.8cm}
\begin{figure}[!htbp]
\includegraphics[width=0.9\textwidth]{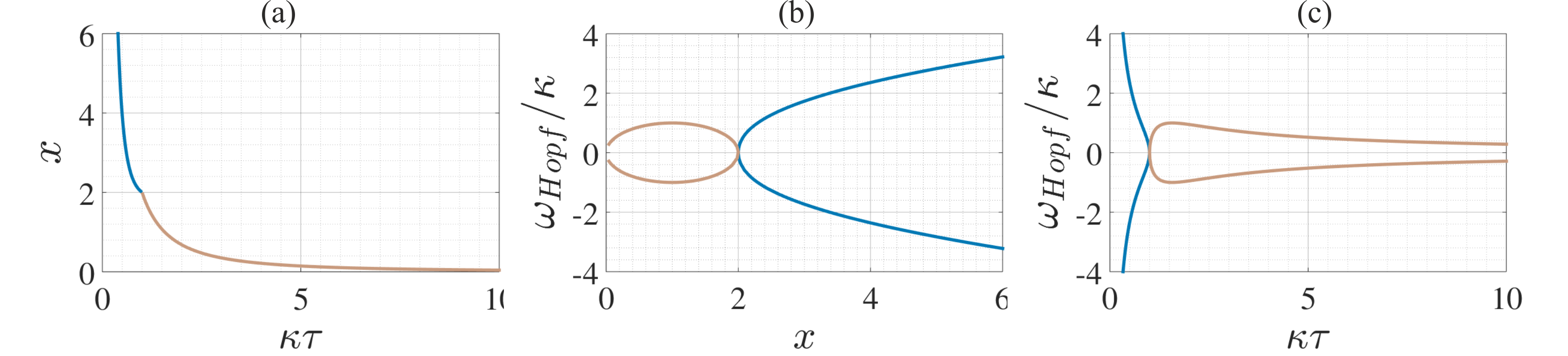}
\vspace{-.3cm}
\caption{Tunability of Hopf bifurcation with (blue) and without pump depletion (brown). (a) Location of Hopf bifurcations in the parameter space described by the pump constant $x$ and the time delay $\tau$. (b,c) Characteristic frequency of oscillations starting above the Hopf bifurcation as a function of the driving strength and time delay. Parameters are $\kappa_p=\kappa$, $k=\kappa$, $\phi=0$.}\label{fig:ddetuning}
\end{figure}
\vspace{-.3cm}

\end{widetext}

\section{Tunability of the frequency of the best squeezing}

In the plant-controller setup of \cite{Crisafulli2013}, tunable sideband squeezing was found in the destructive interference case, which corresponds to our choice $\phi=0$.
According to the results at the end of the previous section, in our setup the Hopf frequency can be tuned by changing, e.g., the driving strength or the time delay. In this section we examine how much the frequency corresponding to perfect squeezing can be varied, when $\phi=0$ and pump depletion is not considered. By using \refeq{spec_tau} one can obtain boundaries for the feedback strength as
\begin{align}
|\kappa-|\epsilon||\le k \le\kappa .
\end{align}
With the help of this relation and \refeq{spec_tau}, the following frequency range can be obtained,
\begin{align}
0\le\nu_c\le\sqrt{|\epsilon|(2\kappa-|\epsilon|)},
\end{align}
which has a maximum of $\nu_c=\kappa$ at $|\epsilon|=k=\kappa$, similar to Fig.~\ref{fig:ddetuning}(b).

Our quantum mechanical considerations are valid only below threshold, with an undepleted pump mode, i.e., in the stable pump strength range defined in Section \refno{sec:stability}.

Tuning properties of the characteristic frequency of perfect squeezing are collected in Fig.~\ref{fig:3D}, where the effect of loss is also demonstrated. The dark blue path in Fig.~\ref{fig:3D}(a) represents perfect squeezing and aligns well with the light brown curve in Fig.~\ref{fig:ddetuning}(a). Meanwhile, the height of the graph gives the same shape as the light brown curves (of positive frequencies) shown as a function of the driving strength in Fig.~\ref{fig:3D}(b) and of the time-delay in Fig.~\ref{fig:3D}(c).

\enlargethispage{1cm}

Note that, according to \refeq{spec_tau}, the characteristic frequency can also be tuned by altering the feedback strength $k$. This quantity can be changed by setting a different ratio for the transmission rates of the mirrors, or by adjusting the loss rate (see \refeq{k}).

\begin{widetext}

\vspace{-.5cm}
\begin{figure}[ht!]
\centering
\includegraphics[width=0.94\textwidth]{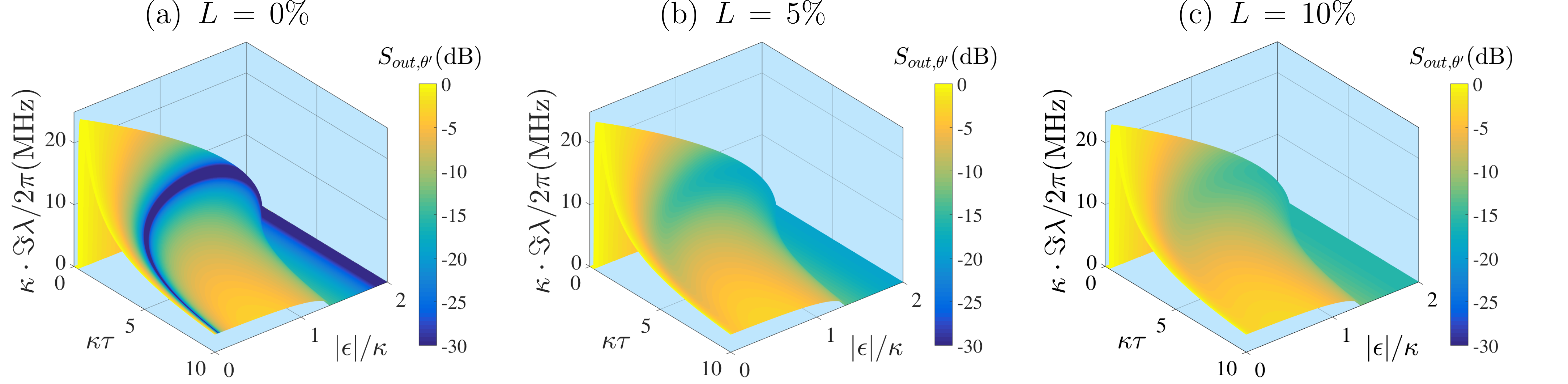}
\vspace{-.3cm}
\caption{Imaginary part of the stability eigenvalue with the highest real part, determined by the equations \refeq{real} and \refeq{imag}, coloured according to the corresponding value of squeezing. Parameters are $\kappa=2\pi\cdot10\text{MHz}$, $\kappa_b=\kappa_c=\kappa/2$, $\Delta=0$, and $\phi=0$.}\label{fig:3D}
\end{figure}
\end{widetext}

\newpage
\section{Influence of detuning and phase shift} \label{sec:detune}

\subsection{Effect of detuning}

To evaluate the effect of a finite detuning $\Delta=\omega_a-\omega_p/2$, we can again look at the squeezing spectrum (\ref{eq:chi_nu}) and, in the ideal case of no loss in the feedback loop, 
determine the optimal quadrature angle for squeezing by setting the expression \refeq{eq:chi_nu} to zero. In particular, for any detuning with $|\Delta | \leq |\epsilon |$, perfect squeezing occurs for 
\begin{align}
\thetap&=\theta + \pi - \arcsin{\lk\frac{\Delta}{|\epsilon|}\rk} ,
\label{spec_freq_delta}
\end{align}
with
\begin{align}
\nu_{c,\Delta}&=\pm\sqrt{k^2-\lk|\epsilon_\Delta|-\kappa\rk^2} , \\
\tau_{c,\Delta}&=\frac{\arccos{\lk\frac{|\epsilon_\Delta|-\kappa}{k}\rk}}{\nu_{c,\Delta}} ,
\end{align}
where
\begin{align}
\label{eps_del}
|\epsilon_\Delta|=\sqrt{|\epsilon|^2-\Delta^2} .
\end{align}

\begin{figure}[!htbp]
\centering
\includegraphics[width=.38\textwidth]{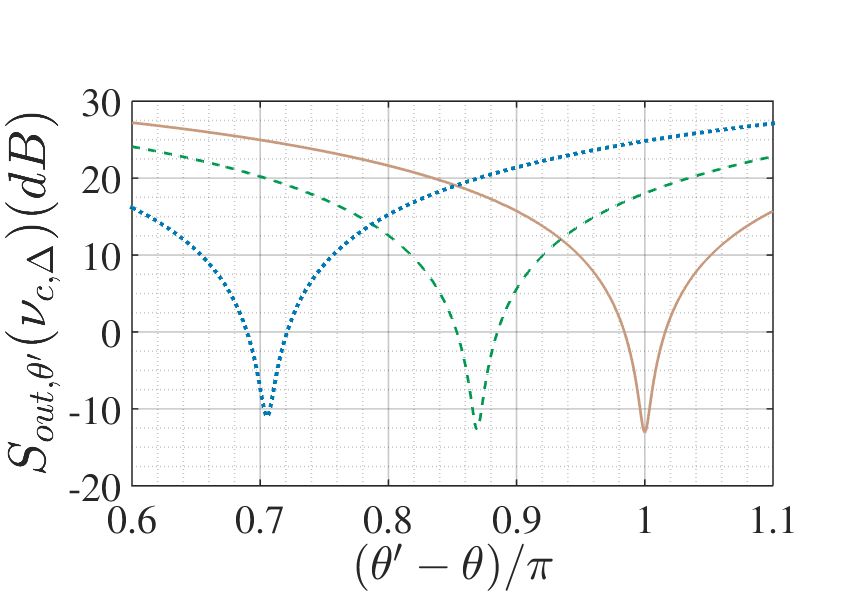}
\vspace{-.3cm}
\caption{Squeezing at $\nu_{c,\Delta}$ and $\tau_{c,\Delta}$ for different values of detuning, $\Delta=0$ (brown solid), $\Delta=0.2\kappa$ (green dashed), and $\Delta=0.4\kappa$ (blue dotted) as a function of the quadrature angle $\thetap$. Parameters are $|\epsilon|=\kappa/2$, $\kappa_b=\kappa_c=\kappa/2$, $\phi=0$, $L=0.05$.}\label{fig:delta_quadr}
\end{figure}

Fig.~\ref{fig:delta_quadr} shows the best possible squeezing obtained, with finite loss ($L>0$), at $\nu=\nu_{c,\Delta}$ and $\tau=\tau_{c,\Delta}$, as a function of quadrature phase angle, for feedback-loop phase $\phi=0$ and several values of $\Delta <|\epsilon|$. The shift in phase angle of the optimal squeezing is well described by (\ref{spec_freq_delta}), while the degree of squeezing is still given by \refeq{Soutpi}.

\subsection{Effect of feedback-loop phase shift}

\begin{figure}[!htbp]
\centering
\includegraphics[width=.38\textwidth]{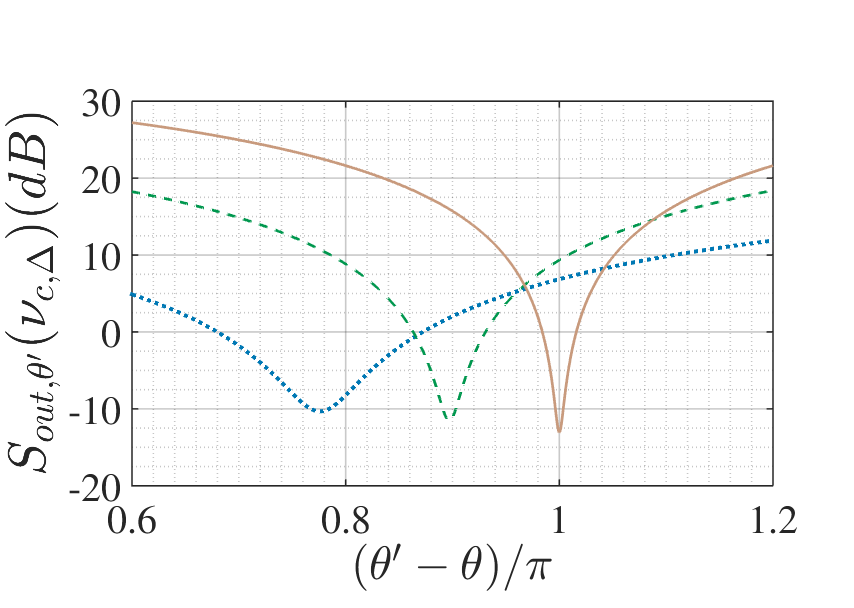}
\vspace{-.3cm}
\caption{Squeezing at $\nu_{c}$ (as determined for $\phi=0$) as a function of the quadrature angle $\thetap$ for different values of feedback-loop phase shift, $\phi=0$ (brown solid), $\phi=-0.05\pi$ (green dashed), and $\phi=-0.1\pi$ (blue dotted). Other parameters are $|\epsilon|=\kappa/2$, $\kappa_b=\kappa_c=\kappa/2$, $\kappa\tau =2.4358$, $\Delta=0$, and $L=0.05$.}\label{fig:phi_quadr}
\end{figure}

Besides influencing the effective feedback strength, the overall feedback-loop phase shift, $\phi$, has a similar effect on the squeezing as the detuning of the subharmonic mode. This is illustrated in Fig.~\ref{fig:phi_quadr}, where squeezing at the frequency $\nu_c$, as determined for $\phi=0$, is shown as a function of quadrature phase angle for $\phi=0$ and two other values of $\phi$ shifted slightly from zero. On top of a shift of the optimal quadrature phase angle, we note that the degree of best squeezing at $\nu_c$ is reduced somewhat with finite $\phi$, although not substantially.

\subsection{Frequency-dependent optimal-squeezing quadrature angle}

\begin{figure}[!htbp]
\centering
\includegraphics[width=0.42\textwidth]{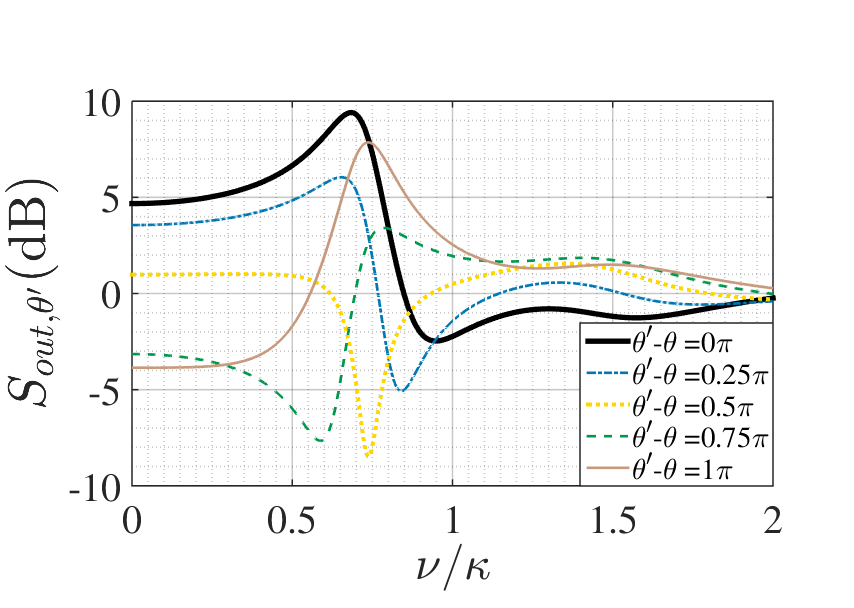}
\vspace{-.3cm}
\caption{Spectrum of squeezing for varying quadrature phase angle, $\theta_d=\thetap-\theta$. Parameters are $|\epsilon|=\kappa/2$, $\kappa_b=0.3\kappa$, $\kappa\tau=2$, $\Delta=0$, $\phi = -0.3\pi$, $L=5\%$.}\label{fig:gravwaves}
\end{figure}

With more general combinations of parameters, more complex variation of the squeezing spectra can be obtained. For example, as illustrated in Fig.~\ref{fig:gravwaves}, it is possible to find parameters for which the frequency of best squeezing varies with quadrature phase angle. In particular, for the example shown, the best squeezing shifts monotonically from resonance ($\nu/\kappa =0$) to higher frequency ($\nu/\kappa\simeq 1$) as the quadrature phase angle increases from zero to $\pi$ radians. 

We note that in the interferometric setups used for gravitational wave detection, in order to reduce the noise in the mechanical mode induced by the amplitude fluctuations in the radiation pressure, amplitude squeezed light is desirable at lower frequencies, whereas for higher frequencies, reduction in the phase quadrature variance is more important in the input channel \cite{Kimble2002}. The example in Fig.~\ref{fig:gravwaves} is potentially of some interest in this context, but requires further investigation that is outside the scope of the present work.

\vspace{-.5cm}

\section{Pyragas-type feedback}\label{sec:Pyragas}

In this section, we consider in more detail the case with feedback-loop phase shift $\phi=\pi$, where it becomes possible to rewrite the equation of motion for the intracavity field operator in the form
\begin{align}
\tder{\aop(t)}=&i[\Hop,\aop(t)]-(\kappa-k)\aop(t)-\sqrt{2\kappa}\,\aop_{in}(t) \nn
\\
& + k\lk\aop(t-\tau)-\aop(t)\rk .
\end{align}
This, in fact, has the characteristic form of a (classical) feedback scheme originally introduced by Pyragas to stabilize unstable periodic orbits of chaotic dynamical systems by feeding back the output signal with a time delay adjusted to the desired period \cite{Pyragas1992}. Feedback of this form has recently been considered in the context of quantum-optical systems, with, in particular, potential for faster convergence to steady states demonstrated \cite{Grimsmo2014,Naumann2014}. 

\begin{figure}[!htbp]
\centering
\includegraphics[width=0.42\textwidth]{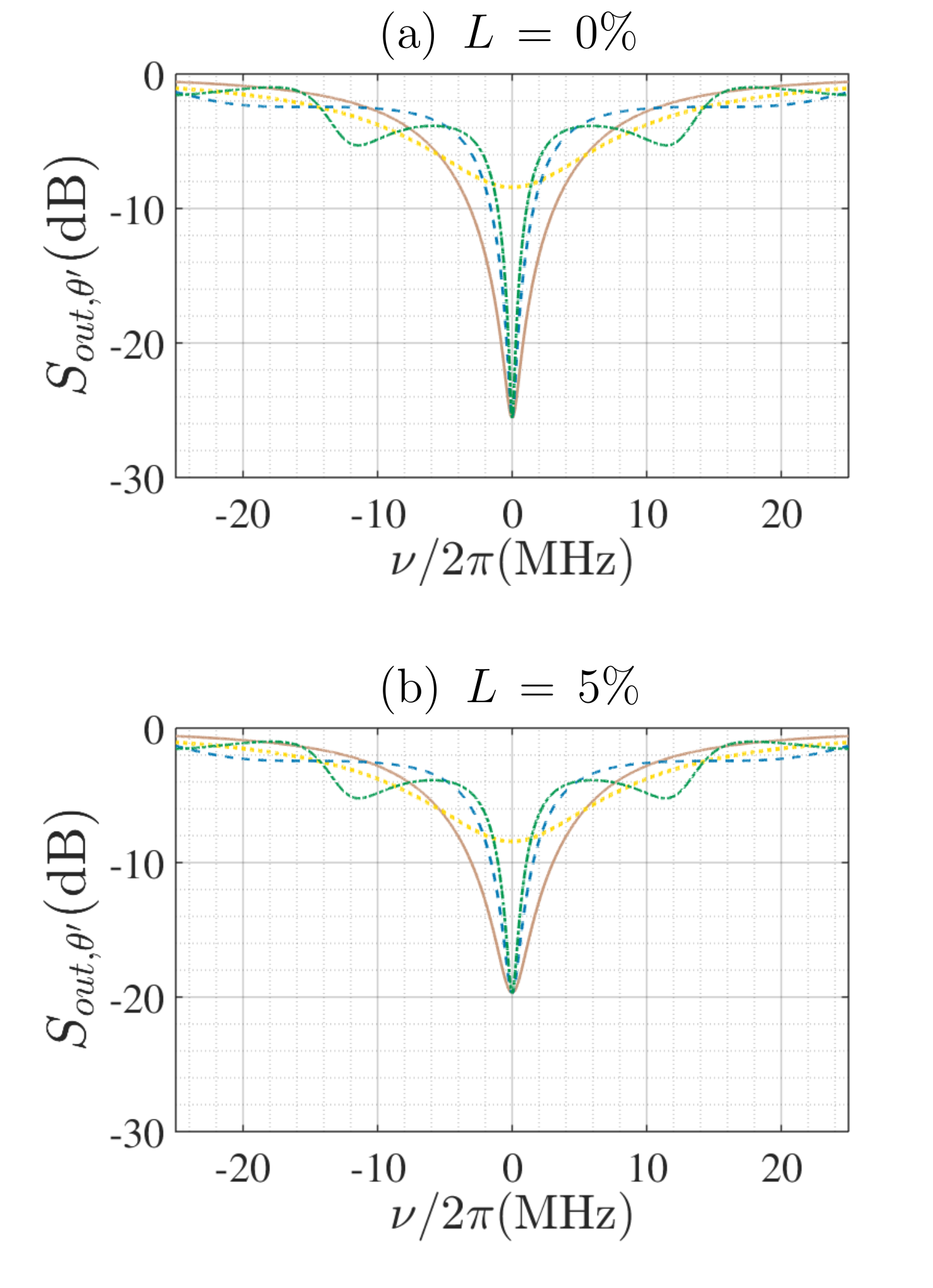}
\vspace{-.3cm}\caption{Squeezing spectra at different values of time-delay. Theoretical results without feedback for the one-sided case in yellow dotted \cite{Collett1984}, and our results with feedback in brown solid ($\kappa\tau=0$), blue dashed ($\kappa\tau=2$) and green dot-dashed ($\kappa\tau=4$). Other parameters are $\thetap=\theta+\pi$, $\kappa=2\pi\cdot10 \text{ MHz}$, $|\epsilon|=0.45\kappa$, $k=\frac{\kappa}{2}\sqrt{1-L}$, $\Delta=0$, and $\phi=\pi$.}\label{fig:squeez_Pyragas}
\end{figure}

As we have already shown in Section~\refno{sec:stab_pyrag}, the system with $\phi=\pi$ is stable as long as $\kappa-k>|\epsilon|$. A consequence of this is that the amount of time delay does not influence the stability of the steady state solutions below threshold. In Fig.~\ref{fig:squeez_Pyragas} examples of the squeezing spectra are shown with and without loss in the feedback loop. We consider the case $k=(\kappa/2)\sqrt{1-L}$,which corresponds to a cavity that is almost one-sided ($\kappa_b=0.933\kappa$), and is such that, when $L=0$, we have $\kappa-k=k$. 
The effect of the feedback is pronounced, giving rise to significantly enhanced squeezing around resonance. While for longer time delays side-peaks develop in the spectrum, the primary effect of time delay is restricted to a narrowing of the effective bandwidth of squeezing around resonance (similar to \cite{Iida2012}), where the squeezing also remains highest.

\begin{figure}[!htbp]
\centering
\includegraphics[width =0.38\textwidth]{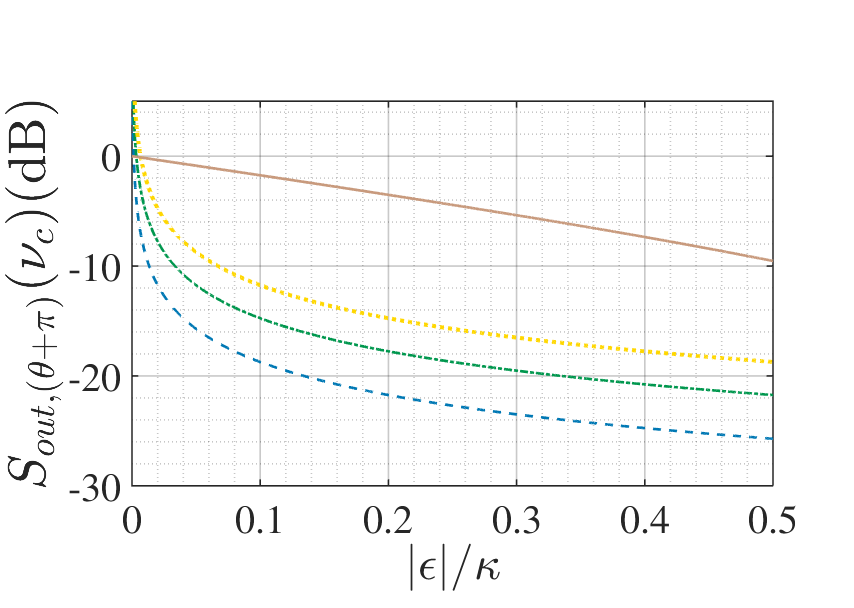}
\vspace{-.3cm}\caption{Maximum squeezing obtained in our model with losses $L=0.02$ (blue dashed), $L=0.05$ (green dot-dashed), $L=0.1$ (yellow dotted), compared to the theoretical limit of squeezing of a DPA in a one-sided cavity (brown-solid). Parameters for our system are $k=\frac{\kappa}{2}\sqrt{1-L}$, $\Delta=0$, $\phi=\pi$, and $\nu_c=0$.}\label{fig:best_Pyragas}
\end{figure}

Focussing on the squeezing at resonance with a finite loss $L$, we find that the enhancement over the no-feedback system is significant over the whole, stable range of pump strengths, as demonstrated in Fig.~\refno{fig:best_Pyragas}. Even with 10\% loss ($L=0.1$), the potential squeezing is 10~dB more than what is anticipated in an ideal one-sided cavity without feedback.

\begin{figure}[!htbp]
\centering
\includegraphics[width=0.38\textwidth]{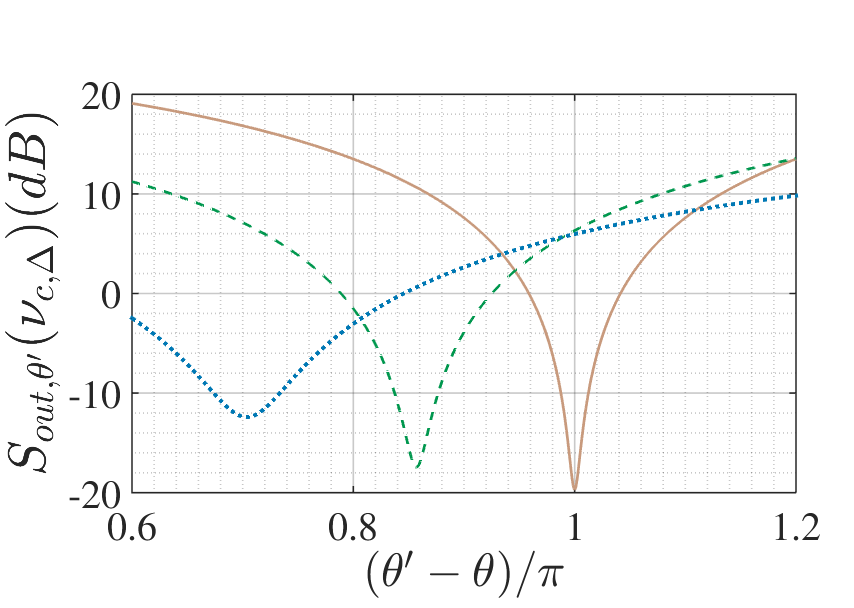}
\vspace{-.3cm}\caption{Squeezing at $\nu=0$ for $\phi=\pi$ and different values of detuning, $\Delta=0$ (brown solid), $\Delta=0.2\kappa$ (green dashed), and $\Delta=0.4\kappa$ (blue dotted), as a function of the quadrature angle $\thetap$. Other parameters are $|\epsilon|=0.45\kappa$, $k=\frac{\kappa}{2}\sqrt{1-L}$, $\tau=0$, and $L=0.05$.}\label{fig:squeez_Pyragas_det}
\end{figure}

\begin{figure}[!htbp]
\centering
\includegraphics[width=0.38\textwidth]{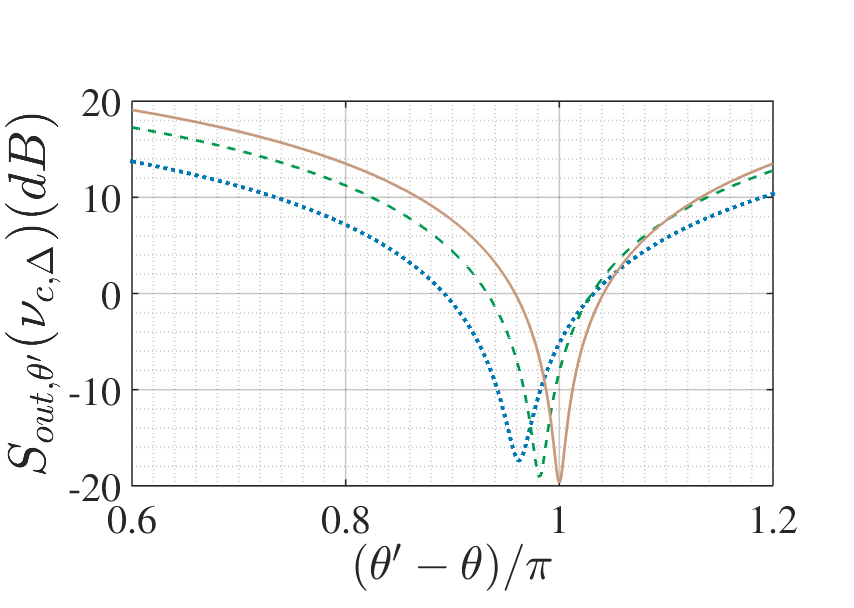}
\vspace{-.3cm}\caption{Squeezing at $\nu=0$ for $\Delta=0$ and different values of the feedback-loop phase shift, $\phi=\pi$ (brown solid), $\phi=0.95\pi$ (green dashed), and $\phi=0.9\pi$ (blue dotted), as a function of the quadrature angle $\thetap$. Other parameters are $|\epsilon|=0.45\kappa$, $k=\frac{\kappa}{2}\sqrt{1-L}$, $\tau=0$, and $L=0.05$.}\label{fig:squeez_Pyragas_phi}
\end{figure}

To round out our examination of this Pyragas-type feedback, and contrast with the case of $\phi=0$, in Figs.~\ref{fig:squeez_Pyragas_det} and \ref{fig:squeez_Pyragas_phi} we briefly examine the effect of detuning and of variations in feedback-loop phase shift, respectively. Not surprisingly, finite detuning diminishes the squeezing on resonance in this setup, while also shifting the optimal quadrature phase angle. Similarly, small shifts in the feedback-loop phase away from $\phi=\pi$ decrease the best squeezing, albeit only slightly, and cause a small shift in the optimal quadrature phase angle (notably smaller, however, than the shift induced by the corresponding deviations from the $\phi=0$ case, as shown in Fig.~\ref{fig:phi_quadr}).

\section{Conclusion}

In this paper we have investigated the influence of time-delayed coherent feedback on the output squeezing of a degenerate parametric amplifier. While some of our findings reproduce, qualitatively, certain results of previous feedback studies \cite{Gough2009,Bian2012,Iida2012,Crisafulli2013}, we also find new and promising operating conditions that should in principle enable significantly enhanced squeezing either around the cavity resonance ($\phi=\pi$), or in sidebands detuned by up to the cavity linewidth from the cavity resonance ($\phi=0$). 

Enhanced squeezing around resonance arises for a particular choice of feedback-loop phase shift, $\phi=\pi$, that gives rise to Pyragas-type feedback and corresponds to an effective reduction in the threshold pump power. Substantial enhancement persists even with significant loss in the feedback loop. Alternatively, with feedback-loop phase shift $\phi=0$, substantial enhancement is possible in sidebands detuned from resonance, provided there is a time delay in the feedback loop on the order of the cavity lifetime. In both cases, the enhancement is also found to be robust to finite cavity detunings and small changes in feedback-loop phase shift.

In order to help interpret the obtained results, a classical analysis of the dynamics was also performed. Perfect squeezing was demonstrated to be related to the classical system's changing stability and the emerging internal dynamics as described by the Hopf frequency. 

Given these novel and promising results obtained with a simple degenerate parametric amplifier, it will be interesting to consider the influence of similar, time-delayed coherent feedback on the quantum mechanical properties of related quantum optical systems, such as the non-degenerate parametric amplifier and cavity-optomechanical configurations.  

Note: Near completion of this manuscript, we learned of closely related work by M. Kraft {\em et al}. \cite{Kraft2016}, who consider a time-delayed coherent feedback loop applied to one side of two-sided DPA.

\acknowledgments
The authors thank Howard Carmichael for discussions and acknowledge support from the Marsden Fund of the Royal Society of New Zealand. SP gratefully acknowledges the IQIM and the group of Jeff Kimble at Caltech for support and hospitality during a visit when this work was initiated.

\bibliographystyle{apsrev4-1}
\bibliography{DPA_squeez_paper2}

\begin{thebibliography}{31}%
\makeatletter
\providecommand \@ifxundefined [1]{%
 \@ifx{#1\undefined}
}%
\providecommand \@ifnum [1]{%
 \ifnum #1\expandafter \@firstoftwo
 \else \expandafter \@secondoftwo
 \fi
}%
\providecommand \@ifx [1]{%
 \ifx #1\expandafter \@firstoftwo
 \else \expandafter \@secondoftwo
 \fi
}%
\providecommand \natexlab [1]{#1}%
\providecommand \enquote  [1]{``#1''}%
\providecommand \bibnamefont  [1]{#1}%
\providecommand \bibfnamefont [1]{#1}%
\providecommand \citenamefont [1]{#1}%
\providecommand \href@noop [0]{\@secondoftwo}%
\providecommand \href [0]{\begingroup \@sanitize@url \@href}%
\providecommand \@href[1]{\@@startlink{#1}\@@href}%
\providecommand \@@href[1]{\endgroup#1\@@endlink}%
\providecommand \@sanitize@url [0]{\catcode `\\12\catcode `\$12\catcode
  `\&12\catcode `\#12\catcode `\^12\catcode `\_12\catcode `\%12\relax}%
\providecommand \@@startlink[1]{}%
\providecommand \@@endlink[0]{}%
\providecommand \url  [0]{\begingroup\@sanitize@url \@url }%
\providecommand \@url [1]{\endgroup\@href {#1}{\urlprefix }}%
\providecommand \urlprefix  [0]{URL }%
\providecommand \Eprint [0]{\href }%
\providecommand \doibase [0]{http://dx.doi.org/}%
\providecommand \selectlanguage [0]{\@gobble}%
\providecommand \bibinfo  [0]{\@secondoftwo}%
\providecommand \bibfield  [0]{\@secondoftwo}%
\providecommand \translation [1]{[#1]}%
\providecommand \BibitemOpen [0]{}%
\providecommand \bibitemStop [0]{}%
\providecommand \bibitemNoStop [0]{.\EOS\space}%
\providecommand \EOS [0]{\spacefactor3000\relax}%
\providecommand \BibitemShut  [1]{\csname bibitem#1\endcsname}%
\let\auto@bib@innerbib\@empty
\bibitem [{\citenamefont {Wisemann}\ and\ \citenamefont
  {Milburn}(1994)}]{Wisemann1994}%
  \BibitemOpen
  \bibfield  {author} {\bibinfo {author} {\bibfnamefont {H.~M.}\ \bibnamefont
  {Wisemann}}\ and\ \bibinfo {author} {\bibfnamefont {G.~J.}\ \bibnamefont
  {Milburn}},\ }\href {\doibase 10.1103/PhysRevA.49.1350} {\bibfield  {journal}
  {\bibinfo  {journal} {Phys. Rev. A}\ }\textbf {\bibinfo {volume} {49}},\
  \bibinfo {pages} {1350} (\bibinfo {year} {1994})}\BibitemShut {NoStop}%
\bibitem [{\citenamefont {Lloyd}(2000)}]{Lloyd2000}%
  \BibitemOpen
  \bibfield  {author} {\bibinfo {author} {\bibfnamefont {S.}~\bibnamefont
  {Lloyd}},\ }\href {\doibase 10.1103/PhysRevA.62.022108} {\bibfield  {journal}
  {\bibinfo  {journal} {Phys. Rev. A}\ }\textbf {\bibinfo {volume} {62}},\
  \bibinfo {pages} {022108} (\bibinfo {year} {2000})}\BibitemShut {NoStop}%
\bibitem [{\citenamefont {Nelson}\ \emph {et~al.}(2000)\citenamefont {Nelson},
  \citenamefont {Weinstein}, \citenamefont {Cory},\ and\ \citenamefont
  {Lloyd}}]{Nelson2000}%
  \BibitemOpen
  \bibfield  {author} {\bibinfo {author} {\bibfnamefont {R.~J.}\ \bibnamefont
  {Nelson}}, \bibinfo {author} {\bibfnamefont {Y.}~\bibnamefont {Weinstein}},
  \bibinfo {author} {\bibfnamefont {D.}~\bibnamefont {Cory}}, \ and\ \bibinfo
  {author} {\bibfnamefont {S.}~\bibnamefont {Lloyd}},\ }\href {\doibase
  10.1103/PhysRevLett.85.3045} {\bibfield  {journal} {\bibinfo  {journal}
  {Phys. Rev. Lett.}\ }\textbf {\bibinfo {volume} {85}},\ \bibinfo {pages}
  {3045} (\bibinfo {year} {2000})}\BibitemShut {NoStop}%
\bibitem [{\citenamefont {Mabuchi}(2008)}]{Mabuchi2008a}%
  \BibitemOpen
  \bibfield  {author} {\bibinfo {author} {\bibfnamefont {H.}~\bibnamefont
  {Mabuchi}},\ }\href {\doibase 10.1103/PhysRevA.78.032323} {\bibfield
  {journal} {\bibinfo  {journal} {Phys. Rev. A}\ }\textbf {\bibinfo {volume}
  {78}},\ \bibinfo {pages} {032323} (\bibinfo {year} {2008})},\ \Eprint
  {http://arxiv.org/abs/0803.2007} {arXiv:0803.2007} \BibitemShut {NoStop}%
\bibitem [{\citenamefont {Gough}\ and\ \citenamefont
  {Wildfeuer}(2009)}]{Gough2009}%
  \BibitemOpen
  \bibfield  {author} {\bibinfo {author} {\bibfnamefont {J.~E.}\ \bibnamefont
  {Gough}}\ and\ \bibinfo {author} {\bibfnamefont {S.}~\bibnamefont
  {Wildfeuer}},\ }\href@noop {} {\bibfield  {journal} {\bibinfo  {journal}
  {Phys. Rev. A}\ }\textbf {\bibinfo {volume} {80}},\ \bibinfo {pages} {042107}
  (\bibinfo {year} {2009})}\BibitemShut {NoStop}%
\bibitem [{\citenamefont {Yan}\ \emph {et~al.}(2011)\citenamefont {Yan},
  \citenamefont {Jia}, \citenamefont {Xie},\ and\ \citenamefont
  {Peng}}]{Yan2011}%
  \BibitemOpen
  \bibfield  {author} {\bibinfo {author} {\bibfnamefont {Z.}~\bibnamefont
  {Yan}}, \bibinfo {author} {\bibfnamefont {X.}~\bibnamefont {Jia}}, \bibinfo
  {author} {\bibfnamefont {C.}~\bibnamefont {Xie}}, \ and\ \bibinfo {author}
  {\bibfnamefont {K.}~\bibnamefont {Peng}},\ }\href {\doibase
  10.1103/PhysRevA.84.062304} {\bibfield  {journal} {\bibinfo  {journal} {Phys.
  Rev. A}\ }\textbf {\bibinfo {volume} {84}},\ \bibinfo {pages} {062304}
  (\bibinfo {year} {2011})},\ \Eprint {http://arxiv.org/abs/1201.1163v1}
  {arXiv:1201.1163v1} \BibitemShut {NoStop}%
\bibitem [{\citenamefont {Iida}\ \emph {et~al.}(2012)\citenamefont {Iida},
  \citenamefont {Yukawa}, \citenamefont {Yonezawa}, \citenamefont {Yamamoto},\
  and\ \citenamefont {Furusawa}}]{Iida2012}%
  \BibitemOpen
  \bibfield  {author} {\bibinfo {author} {\bibfnamefont {S.}~\bibnamefont
  {Iida}}, \bibinfo {author} {\bibfnamefont {M.}~\bibnamefont {Yukawa}},
  \bibinfo {author} {\bibfnamefont {H.}~\bibnamefont {Yonezawa}}, \bibinfo
  {author} {\bibfnamefont {N.}~\bibnamefont {Yamamoto}}, \ and\ \bibinfo
  {author} {\bibfnamefont {A.}~\bibnamefont {Furusawa}},\ }\href@noop {}
  {\bibfield  {journal} {\bibinfo  {journal} {IEEE Trans. Automat. Contr.}\
  }\textbf {\bibinfo {volume} {57}},\ \bibinfo {pages} {2045} (\bibinfo {year}
  {2012})},\ \Eprint {http://arxiv.org/abs/1103.1324} {arXiv:1103.1324}
  \BibitemShut {NoStop}%
\bibitem [{\citenamefont {Crisafulli}\ \emph {et~al.}(2013)\citenamefont
  {Crisafulli}, \citenamefont {Tezak}, \citenamefont {Soh}, \citenamefont
  {Armen},\ and\ \citenamefont {Mabuchi}}]{Crisafulli2013}%
  \BibitemOpen
  \bibfield  {author} {\bibinfo {author} {\bibfnamefont {O.}~\bibnamefont
  {Crisafulli}}, \bibinfo {author} {\bibfnamefont {N.}~\bibnamefont {Tezak}},
  \bibinfo {author} {\bibfnamefont {D.~B.~S.}\ \bibnamefont {Soh}}, \bibinfo
  {author} {\bibfnamefont {M.~a.}\ \bibnamefont {Armen}}, \ and\ \bibinfo
  {author} {\bibfnamefont {H.}~\bibnamefont {Mabuchi}},\ }\href {\doibase
  10.1364/OE.21.018371} {\bibfield  {journal} {\bibinfo  {journal} {Opt.
  Express}\ }\textbf {\bibinfo {volume} {21}},\ \bibinfo {pages} {1} (\bibinfo
  {year} {2013})},\ \Eprint {http://arxiv.org/abs/1302.6179} {arXiv:1302.6179}
  \BibitemShut {NoStop}%
\bibitem [{\citenamefont {Jacobs}\ \emph {et~al.}(2014)\citenamefont {Jacobs},
  \citenamefont {Wang},\ and\ \citenamefont {Wiseman}}]{Jacobs2014}%
  \BibitemOpen
  \bibfield  {author} {\bibinfo {author} {\bibfnamefont {K.}~\bibnamefont
  {Jacobs}}, \bibinfo {author} {\bibfnamefont {X.}~\bibnamefont {Wang}}, \ and\
  \bibinfo {author} {\bibfnamefont {H.~M.}\ \bibnamefont {Wiseman}},\ }\href
  {\doibase 10.1088/1367-2630/16/7/073036} {\bibfield  {journal} {\bibinfo
  {journal} {New J. Phys.}\ }\textbf {\bibinfo {volume} {16}},\ \bibinfo
  {pages} {073036} (\bibinfo {year} {2014})},\ \Eprint
  {http://arxiv.org/abs/1211.1724} {arXiv:1211.1724} \BibitemShut {NoStop}%
\bibitem [{\citenamefont {Zhou}\ \emph {et~al.}(2015)\citenamefont {Zhou},
  \citenamefont {Jia}, \citenamefont {Li}, \citenamefont {Yu}, \citenamefont
  {Xie},\ and\ \citenamefont {Peng}}]{Zhou2015}%
  \BibitemOpen
  \bibfield  {author} {\bibinfo {author} {\bibfnamefont {Y.}~\bibnamefont
  {Zhou}}, \bibinfo {author} {\bibfnamefont {X.}~\bibnamefont {Jia}}, \bibinfo
  {author} {\bibfnamefont {F.}~\bibnamefont {Li}}, \bibinfo {author}
  {\bibfnamefont {J.}~\bibnamefont {Yu}}, \bibinfo {author} {\bibfnamefont
  {C.}~\bibnamefont {Xie}}, \ and\ \bibinfo {author} {\bibfnamefont
  {K.}~\bibnamefont {Peng}},\ }\href {\doibase 10.1038/srep11132} {\bibfield
  {journal} {\bibinfo  {journal} {Sci. Rep.}\ }\textbf {\bibinfo {volume}
  {5}},\ \bibinfo {pages} {11132} (\bibinfo {year} {2015})}\BibitemShut
  {NoStop}%
\bibitem [{\citenamefont {Bian}\ \emph {et~al.}(2012)\citenamefont {Bian},
  \citenamefont {Zhang},\ and\ \citenamefont {Lee}}]{Bian2012}%
  \BibitemOpen
  \bibfield  {author} {\bibinfo {author} {\bibfnamefont {C.}~\bibnamefont
  {Bian}}, \bibinfo {author} {\bibfnamefont {G.}~\bibnamefont {Zhang}}, \ and\
  \bibinfo {author} {\bibfnamefont {H.~W.~J.}\ \bibnamefont {Lee}},\ }\href
  {\doibase 10.1080/00207179.2012.706872} {\bibfield  {journal} {\bibinfo
  {journal} {Int. J. Control}\ }\textbf {\bibinfo {volume} {85}},\ \bibinfo
  {pages} {1865} (\bibinfo {year} {2012})}\BibitemShut {NoStop}%
\bibitem [{\citenamefont {{Pyragas K.}}(1992)}]{Pyragas1992}%
  \BibitemOpen
  \bibfield  {author} {\bibinfo {author} {\bibnamefont {{Pyragas K.}}},\ }\href
  {\doibase 10.1016/0375-9601(92)90745-8} {\bibfield  {journal} {\bibinfo
  {journal} {Phys. Lett. A}\ }\textbf {\bibinfo {volume} {170}},\ \bibinfo
  {pages} {421} (\bibinfo {year} {1992})}\BibitemShut {NoStop}%
\bibitem [{\citenamefont {H{\"{o}}vel}\ and\ \citenamefont
  {Sch{\"{o}}ll}(2005)}]{Hovel2005}%
  \BibitemOpen
  \bibfield  {author} {\bibinfo {author} {\bibfnamefont {P.}~\bibnamefont
  {H{\"{o}}vel}}\ and\ \bibinfo {author} {\bibfnamefont {E.}~\bibnamefont
  {Sch{\"{o}}ll}},\ }\href {\doibase 10.1109/PHYCON.2005.1514008} {\bibfield
  {journal} {\bibinfo  {journal} {Phys. Rev. E}\ }\textbf {\bibinfo {volume}
  {72}},\ \bibinfo {pages} {046203} (\bibinfo {year} {2005})},\ \Eprint
  {http://arxiv.org/abs/0508367} {arXiv:0508367} \BibitemShut {NoStop}%
\bibitem [{\citenamefont {Carmele}\ \emph {et~al.}(2013)\citenamefont
  {Carmele}, \citenamefont {Kabuss}, \citenamefont {Schulze}, \citenamefont
  {Reitzenstein},\ and\ \citenamefont {Knorr}}]{Carmele2013}%
  \BibitemOpen
  \bibfield  {author} {\bibinfo {author} {\bibfnamefont {A.}~\bibnamefont
  {Carmele}}, \bibinfo {author} {\bibfnamefont {J.}~\bibnamefont {Kabuss}},
  \bibinfo {author} {\bibfnamefont {F.}~\bibnamefont {Schulze}}, \bibinfo
  {author} {\bibfnamefont {S.}~\bibnamefont {Reitzenstein}}, \ and\ \bibinfo
  {author} {\bibfnamefont {A.}~\bibnamefont {Knorr}},\ }\href {\doibase
  10.1103/PhysRevLett.110.013601} {\bibfield  {journal} {\bibinfo  {journal}
  {Phys. Rev. Lett.}\ }\textbf {\bibinfo {volume} {110}},\ \bibinfo {pages}
  {013601} (\bibinfo {year} {2013})}\BibitemShut {NoStop}%
\bibitem [{\citenamefont {Zhang}\ \emph {et~al.}(2013)\citenamefont {Zhang},
  \citenamefont {Liu}, \citenamefont {Wu}, \citenamefont {Jacobs},\ and\
  \citenamefont {Nori}}]{Zhang2013}%
  \BibitemOpen
  \bibfield  {author} {\bibinfo {author} {\bibfnamefont {J.}~\bibnamefont
  {Zhang}}, \bibinfo {author} {\bibfnamefont {Y.~X.}\ \bibnamefont {Liu}},
  \bibinfo {author} {\bibfnamefont {R.~B.}\ \bibnamefont {Wu}}, \bibinfo
  {author} {\bibfnamefont {K.}~\bibnamefont {Jacobs}}, \ and\ \bibinfo {author}
  {\bibfnamefont {F.}~\bibnamefont {Nori}},\ }\href {\doibase
  10.1103/PhysRevA.49.1350} {\bibfield  {journal} {\bibinfo  {journal} {Phys.
  Rev. A}\ }\textbf {\bibinfo {volume} {87}},\ \bibinfo {pages} {032117}
  (\bibinfo {year} {2013})},\ \Eprint {http://arxiv.org/abs/1208.4720v1}
  {arXiv:1208.4720v1} \BibitemShut {NoStop}%
\bibitem [{\citenamefont {Grimsmo}\ \emph {et~al.}(2014)\citenamefont
  {Grimsmo}, \citenamefont {Parkins},\ and\ \citenamefont
  {Skagerstam}}]{Grimsmo2014}%
  \BibitemOpen
  \bibfield  {author} {\bibinfo {author} {\bibfnamefont {A.~L.}\ \bibnamefont
  {Grimsmo}}, \bibinfo {author} {\bibfnamefont {A.~S.}\ \bibnamefont
  {Parkins}}, \ and\ \bibinfo {author} {\bibfnamefont {B.~S.}\ \bibnamefont
  {Skagerstam}},\ }\href {\doibase 10.1088/1367-2630/16/6/065004} {\bibfield
  {journal} {\bibinfo  {journal} {New J. Phys.}\ }\textbf {\bibinfo {volume}
  {16}},\ \bibinfo {pages} {065004} (\bibinfo {year} {2014})},\ \Eprint
  {http://arxiv.org/abs/1401.2287} {arXiv:1401.2287} \BibitemShut {NoStop}%
\bibitem [{\citenamefont {Naumann}\ \emph {et~al.}(2014)\citenamefont
  {Naumann}, \citenamefont {Hein}, \citenamefont {Knorr},\ and\ \citenamefont
  {Kabuss}}]{Naumann2014}%
  \BibitemOpen
  \bibfield  {author} {\bibinfo {author} {\bibfnamefont {N.~L.}\ \bibnamefont
  {Naumann}}, \bibinfo {author} {\bibfnamefont {S.~M.}\ \bibnamefont {Hein}},
  \bibinfo {author} {\bibfnamefont {A.}~\bibnamefont {Knorr}}, \ and\ \bibinfo
  {author} {\bibfnamefont {J.}~\bibnamefont {Kabuss}},\ }\href {\doibase
  10.1103/PhysRevA.90.043835} {\bibfield  {journal} {\bibinfo  {journal} {Phys.
  Rev. A}\ }\textbf {\bibinfo {volume} {90}},\ \bibinfo {pages} {043835}
  (\bibinfo {year} {2014})}\BibitemShut {NoStop}%
\bibitem [{\citenamefont {Kopylov}\ \emph
  {et~al.}(2015{\natexlab{a}})\citenamefont {Kopylov}, \citenamefont
  {Radonji\'{c}}, \citenamefont {Brandes}, \citenamefont {Bala\v{z}},\ and\
  \citenamefont {Pelster}}]{Kopylov2015a}%
  \BibitemOpen
  \bibfield  {author} {\bibinfo {author} {\bibfnamefont {W.}~\bibnamefont
  {Kopylov}}, \bibinfo {author} {\bibfnamefont {M.}~\bibnamefont
  {Radonji\'{c}}}, \bibinfo {author} {\bibfnamefont {T.}~\bibnamefont
  {Brandes}}, \bibinfo {author} {\bibfnamefont {A.}~\bibnamefont {Bala\v{z}}},
  \ and\ \bibinfo {author} {\bibfnamefont {A.}~\bibnamefont {Pelster}},\ }\href
  {\doibase 10.1103/PhysRevA.92.063832} {\bibfield  {journal} {\bibinfo
  {journal} {Phys. Rev. A}\ }\textbf {\bibinfo {volume} {92}},\ \bibinfo
  {pages} {063832} (\bibinfo {year} {2015}{\natexlab{a}})},\ \Eprint
  {http://arxiv.org/abs/1507.01811} {arXiv:1507.01811} \BibitemShut {NoStop}%
\bibitem [{\citenamefont {Kopylov}\ \emph
  {et~al.}(2015{\natexlab{b}})\citenamefont {Kopylov}, \citenamefont {Emary},
  \citenamefont {Sch\"{o}ll},\ and\ \citenamefont {Brandes}}]{Kopylov2015b}%
  \BibitemOpen
  \bibfield  {author} {\bibinfo {author} {\bibfnamefont {W.}~\bibnamefont
  {Kopylov}}, \bibinfo {author} {\bibfnamefont {C.}~\bibnamefont {Emary}},
  \bibinfo {author} {\bibfnamefont {E.}~\bibnamefont {Sch\"{o}ll}}, \ and\
  \bibinfo {author} {\bibfnamefont {T.}~\bibnamefont {Brandes}},\ }\href
  {\doibase 10.1088/1367-2630/17/1/013040} {\bibfield  {journal} {\bibinfo
  {journal} {New J. Phys.}\ }\textbf {\bibinfo {volume} {17}},\ \bibinfo
  {pages} {13040} (\bibinfo {year} {2015}{\natexlab{b}})},\ \Eprint
  {http://arxiv.org/abs/1403.0620v2} {arXiv:1403.0620v2} \BibitemShut {NoStop}%
\bibitem [{\citenamefont {Hein}\ \emph {et~al.}(2015)\citenamefont {Hein},
  \citenamefont {Schulze}, \citenamefont {Carmele},\ and\ \citenamefont
  {Knorr}}]{Hein2015}%
  \BibitemOpen
  \bibfield  {author} {\bibinfo {author} {\bibfnamefont {S.~M.}\ \bibnamefont
  {Hein}}, \bibinfo {author} {\bibfnamefont {F.}~\bibnamefont {Schulze}},
  \bibinfo {author} {\bibfnamefont {A.}~\bibnamefont {Carmele}}, \ and\
  \bibinfo {author} {\bibfnamefont {A.}~\bibnamefont {Knorr}},\ }\href
  {\doibase 10.1103/PhysRevA.91.052321} {\bibfield  {journal} {\bibinfo
  {journal} {Phys. Rev. A}\ }\textbf {\bibinfo {volume} {91}},\ \bibinfo
  {pages} {052321} (\bibinfo {year} {2015})}\BibitemShut {NoStop}%
\bibitem [{\citenamefont {Kabuss}\ \emph {et~al.}(2015)\citenamefont {Kabuss},
  \citenamefont {Krimer}, \citenamefont {Rotter}, \citenamefont {Stannigel},
  \citenamefont {Knorr},\ and\ \citenamefont {Carmele}}]{Kabuss2015}%
  \BibitemOpen
  \bibfield  {author} {\bibinfo {author} {\bibfnamefont {J.}~\bibnamefont
  {Kabuss}}, \bibinfo {author} {\bibfnamefont {D.~O.}\ \bibnamefont {Krimer}},
  \bibinfo {author} {\bibfnamefont {S.}~\bibnamefont {Rotter}}, \bibinfo
  {author} {\bibfnamefont {K.}~\bibnamefont {Stannigel}}, \bibinfo {author}
  {\bibfnamefont {A.}~\bibnamefont {Knorr}}, \ and\ \bibinfo {author}
  {\bibfnamefont {A.}~\bibnamefont {Carmele}},\ }\href {\doibase
  10.1103/PhysRevA.92.053801} {\bibfield  {journal} {\bibinfo  {journal} {Phys.
  Rev. A}\ }\textbf {\bibinfo {volume} {92}},\ \bibinfo {pages} {053801}
  (\bibinfo {year} {2015})}\BibitemShut {NoStop}%
\bibitem [{\citenamefont {Pichler}\ and\ \citenamefont
  {Zoller}(2016)}]{Pichler2016}%
  \BibitemOpen
  \bibfield  {author} {\bibinfo {author} {\bibfnamefont {H.}~\bibnamefont
  {Pichler}}\ and\ \bibinfo {author} {\bibfnamefont {P.}~\bibnamefont
  {Zoller}},\ }\href {\doibase 10.1103/PhysRevLett.116.093601} {\bibfield
  {journal} {\bibinfo  {journal} {Phys. Rev. Lett.}\ }\textbf {\bibinfo
  {volume} {116}},\ \bibinfo {pages} {093601} (\bibinfo {year} {2016})},\
  \Eprint {http://arxiv.org/abs/1510.04646} {arXiv:1510.04646} \BibitemShut
  {NoStop}%
\bibitem [{\citenamefont {Grimsmo}(2015)}]{Grimsmo2015}%
  \BibitemOpen
  \bibfield  {author} {\bibinfo {author} {\bibfnamefont {A.~L.}\ \bibnamefont
  {Grimsmo}},\ }\href {\doibase 10.1103/PhysRevLett.115.060402} {\bibfield
  {journal} {\bibinfo  {journal} {Phys. Rev. Lett.}\ }\textbf {\bibinfo
  {volume} {115}},\ \bibinfo {pages} {060402} (\bibinfo {year} {2015})},\
  \Eprint {http://arxiv.org/abs/1502.06959} {arXiv:1502.06959} \BibitemShut
  {NoStop}%
\bibitem [{\citenamefont {Whalen}\ and\ \citenamefont
  {Carmichael}(2016)}]{Whalen2016}%
  \BibitemOpen
  \bibfield  {author} {\bibinfo {author} {\bibfnamefont {S.~J.}\ \bibnamefont
  {Whalen}}\ and\ \bibinfo {author} {\bibfnamefont {H.~J.}\ \bibnamefont
  {Carmichael}},\ }\href {http://arxiv.org/abs/1602.03971} {\  (\bibinfo {year}
  {2016})},\ \Eprint {http://arxiv.org/abs/1602.03971} {arXiv:1602.03971}
  \BibitemShut {NoStop}%
\bibitem [{\citenamefont {Kabuss}\ \emph {et~al.}(2016)\citenamefont {Kabuss},
  \citenamefont {Katsch}, \citenamefont {Knorr},\ and\ \citenamefont
  {Carmele}}]{Kabuss2016}%
  \BibitemOpen
  \bibfield  {author} {\bibinfo {author} {\bibfnamefont {J.}~\bibnamefont
  {Kabuss}}, \bibinfo {author} {\bibfnamefont {F.}~\bibnamefont {Katsch}},
  \bibinfo {author} {\bibfnamefont {A.}~\bibnamefont {Knorr}}, \ and\ \bibinfo
  {author} {\bibfnamefont {A.}~\bibnamefont {Carmele}},\ }\href {\doibase
  10.1364/JOSAB.33.000C10} {\bibfield  {journal} {\bibinfo  {journal} {J. Opt.
  Soc. Am. B}\ }\textbf {\bibinfo {volume} {33}},\ \bibinfo {pages} {C10}
  (\bibinfo {year} {2016})},\ \Eprint {http://arxiv.org/abs/1512.05884}
  {arXiv:1512.05884} \BibitemShut {NoStop}%
\bibitem [{\citenamefont {Whalen}(2015)}]{Whalen2015}%
  \BibitemOpen
  \bibfield  {author} {\bibinfo {author} {\bibfnamefont {S.}~\bibnamefont
  {Whalen}},\ }\emph {\bibinfo {title} {{Open Quantum Systems with Time-Delayed
  Interactions}}},\ \href@noop {} {Ph.D. thesis},\ \bibinfo  {school}
  {University of Auckland} (\bibinfo {year} {2015})\BibitemShut {NoStop}%
\bibitem [{\citenamefont {Collett}\ and\ \citenamefont
  {Gardiner}(1984)}]{Collett1984}%
  \BibitemOpen
  \bibfield  {author} {\bibinfo {author} {\bibfnamefont {M.~J.}\ \bibnamefont
  {Collett}}\ and\ \bibinfo {author} {\bibfnamefont {C.~W.}\ \bibnamefont
  {Gardiner}},\ }\href {\doibase 10.1103/PhysRevA.30.1386} {\bibfield
  {journal} {\bibinfo  {journal} {Phys. Rev. A}\ }\textbf {\bibinfo {volume}
  {30}},\ \bibinfo {pages} {1386} (\bibinfo {year} {1984})}\BibitemShut
  {NoStop}%
\bibitem [{\citenamefont {Carmichael}(2009)}]{Carmichael2009}%
  \BibitemOpen
  \bibfield  {author} {\bibinfo {author} {\bibfnamefont {H.~J.}\ \bibnamefont
  {Carmichael}},\ }\href {\doibase 10.1007/978-3-540-71320-3} {\emph {\bibinfo
  {title} {{Statistical Methods in Quantum Optics 2: Non-Classical Fields}}}},\
  edited by\ \bibinfo {editor} {\bibfnamefont {W.}~\bibnamefont {Beiglbock}},
  \bibinfo {editor} {\bibfnamefont {J.-P.}\ \bibnamefont {Eckmann}}, \bibinfo
  {editor} {\bibfnamefont {H.}~\bibnamefont {Grosse}}, \bibinfo {editor}
  {\bibfnamefont {M.}~\bibnamefont {Loss}}, \bibinfo {editor} {\bibfnamefont
  {S.}~\bibnamefont {Smirnov}}, \bibinfo {editor} {\bibfnamefont
  {L.}~\bibnamefont {Takhtajan}}, \ and\ \bibinfo {editor} {\bibfnamefont
  {J.}~\bibnamefont {Yngvason}}\ (\bibinfo  {publisher} {Springer},\ \bibinfo
  {year} {2009})\BibitemShut {NoStop}%
\bibitem [{\citenamefont {Tabak}\ and\ \citenamefont
  {Mabuchi}(2016)}]{Tabak2015a}%
  \BibitemOpen
  \bibfield  {author} {\bibinfo {author} {\bibfnamefont {G.}~\bibnamefont
  {Tabak}}\ and\ \bibinfo {author} {\bibfnamefont {H.}~\bibnamefont
  {Mabuchi}},\ }\href {\doibase 10.1140/epjqt/s40507-016-0041-9} {\bibfield
  {journal} {\bibinfo  {journal} {EPJ Quantum Technol.}\ }\textbf {\bibinfo
  {volume} {3}},\ \bibinfo {pages} {3} (\bibinfo {year} {2016})},\ \Eprint
  {http://arxiv.org/abs/1510.08942v1} {arXiv:1510.08942v1} \BibitemShut
  {NoStop}%
\bibitem [{\citenamefont {Kimble}\ \emph {et~al.}(2002)\citenamefont {Kimble},
  \citenamefont {Levin}, \citenamefont {Matsko}, \citenamefont {Thorne},\ and\
  \citenamefont {Vyatchanin}}]{Kimble2002}%
  \BibitemOpen
  \bibfield  {author} {\bibinfo {author} {\bibfnamefont {H.~J.}\ \bibnamefont
  {Kimble}}, \bibinfo {author} {\bibfnamefont {Y.}~\bibnamefont {Levin}},
  \bibinfo {author} {\bibfnamefont {A.~B.}\ \bibnamefont {Matsko}}, \bibinfo
  {author} {\bibfnamefont {K.~S.}\ \bibnamefont {Thorne}}, \ and\ \bibinfo
  {author} {\bibfnamefont {S.~P.}\ \bibnamefont {Vyatchanin}},\ }\href
  {\doibase 10.1103/PhysRevD.65.022002} {\bibfield  {journal} {\bibinfo
  {journal} {Phys. Rev. D}\ }\textbf {\bibinfo {volume} {65}},\ \bibinfo
  {pages} {022002} (\bibinfo {year} {2002})}\BibitemShut {NoStop}%
\bibitem [{\citenamefont {Kraft}\ \emph {et~al.}(2016)\citenamefont {Kraft},
  \citenamefont {Hein}, \citenamefont {Lehnert}, \citenamefont {Sch},
  \citenamefont {Hughes},\ and\ \citenamefont {Knorr}}]{Kraft2016}%
  \BibitemOpen
  \bibfield  {author} {\bibinfo {author} {\bibfnamefont {M.}~\bibnamefont
  {Kraft}}, \bibinfo {author} {\bibfnamefont {S.~M.}\ \bibnamefont {Hein}},
  \bibinfo {author} {\bibfnamefont {J.}~\bibnamefont {Lehnert}}, \bibinfo
  {author} {\bibfnamefont {E.}~\bibnamefont {Sch}}, \bibinfo {author}
  {\bibfnamefont {S.}~\bibnamefont {Hughes}}, \ and\ \bibinfo {author}
  {\bibfnamefont {A.}~\bibnamefont {Knorr}},\ }\href@noop {} {\  (\bibinfo
  {year} {2016})},\ \Eprint {http://arxiv.org/abs/1603.07137v1}
  {arXiv:1603.07137v1} \BibitemShut {NoStop}%
\end{thebibliography}%

\end{document}